\documentclass[%
 reprint,
 amsmath,amssymb,
 aps,
]{revtex4-2}

\usepackage{graphicx}% Include figure files
\usepackage{dcolumn}% Align table columns on decimal point
\usepackage{bm}% bold math
\begin{document}

\preprint{APS/123-QED}

\title{Strong gate-tunability of flat bands in bilayer graphene due to moir\'e encapsulation between hBN monolayers}

\author{Robin Smeyers, Milorad V. Milo\v{s}evi\'c and Lucian Covaci}
\affiliation{Department of Physics and NANOlab Center of Excellence, University of Antwerp, Groenenborgerlaan 171, 2020 Antwerp, Belgium
}

\date{February 2, 2023}
\begin{abstract}
When using hexagonal boron-nitride (hBN) as a substrate for graphene, the resulting moir\'e pattern creates secondary Dirac points. By encapsulating a multilayer graphene within aligned hBN sheets the controlled moir\'e stacking may offer even richer benefits. Using advanced tight-binding simulations on atomistically-relaxed heterostructures, here we show that the gap at the secondary Dirac point can be opened in selected moir\'e-stacking configurations, and is independent of any additional vertical gating of the heterostructure. On the other hand, gating can broadly tune the gap at the principal Dirac point, and may thereby strongly compress the first moir\'e mini-band in width against the moir\'e-induced gap at the secondary Dirac point. We reveal that in hBN-encapsulated bilayer graphene this novel mechanism can lead to isolated bands flatter than 10~meV under moderate gating, hence presenting a convenient pathway towards electronically-controlled strongly-correlated states on demand.
\end{abstract}

\maketitle

\section{Introduction}
The use of hexagonal-boron-nitride (hBN) as a substrate has improved the electronic properties for graphene-based electronic devices when compared to other substrates such as SiO$_2$ \cite{bolotin_ultrahigh_2008, decker_local_2011, dean_boron_2010, xue_scanning_2011, chen_intrinsic_2008}. Similar to graphene, hBN has a hexagonal lattice but with alternating boron and nitrogen atoms instead of carbon. Due to the small lattice mismatch of $\approx 1.8 \% $ between graphene and hBN the stacking configuration changes periodically, in a moir\'e pattern on a scale much larger than the original unit cell. As the stacking changes throughout the moir\'e unit cell, the interatomic hoppings get modulated as well, leading to changes in the electronic properties of graphene such as gaps at the principal Dirac point (PDP), the formation of secondary Dirac points (SDP), and moir\'e minibands \cite{park_new_2008, yankowitz_emergence_2012, wallbank_generic_2013, wallbank_moire_2015, moon_electronic_2014,cea_band_2020, jung_moire_2017}. Due to hBN having a large bandgap, the spectrum of graphene dominates at low energies near the Dirac-point, and the effect of the hBN layers can be seen as an effective periodic potential applied onto the adjacent graphene layer. Previous results have shown that highly-correlated states can emerge in moir\'e systems such as twisted bilayer graphene \cite{cao_unconventional_2018, cao_correlated_2018}, ABC trilayer graphene on hBN \cite{chittari_gate-tunable_2019, chen_evidence_2019} and twisted trilayer graphene\cite{zhang_correlated_2021}, due to formation of ultra flat bands in the electronic band structure. 
Due to its cubic dispersion, ABC-stacked trilayer graphene has proven to be a convenient platform for the generation of flat bands at low energies when moir\'e potentials are applied through twisting \cite{suarez_morell_flat_2010, zhu_twisted_2020, wu_lattice_2021} or by adding an hBN layer \cite{chittari_gate-tunable_2019, chen_evidence_2019}. AB-stacked bilayer graphene has also exhibited hints of ultra-flat dispersions \cite{marchenko_extremely_2018}, which has the added benefit of bypassing the need for precise tuning of the twist angle. Both ABC trilayer graphene and AB bilayer graphene have no bandgap at the PDP, which is topologically protected as long as spatial inversion symmetry is present. An externally applied vertical electric field can be used to break this symmetry and open a gap while also changing the band structure at low energies. The use of an electronic gate to further flatten the band was required in Refs.~\citenum{chittari_gate-tunable_2019} and \citenum{chen_evidence_2019} in order to achieve strongly correlated states, making their emergence externally controllable - a highly desirable feature for envisaged multifunctional devices. 

Despite those numerous successes in combining moir\'e and electronic degrees of freedom to achieve novel properties in multilayer graphene, the role of encapsulation by hBN was mostly reduced to simply improving the structural and behavioral quality of made devices \cite{mayorov_micrometer-scale_2011, kretinin_electronic_2014}. However, it was recently realized that alignment of encapsulating hBN layers does matter \cite{oka_fractal_2021, andelkovic_double_2020, wang_new_2019, long_accurate_2021, kuiri_enhanced_2021}, and that even slight misalignment can induce a strong spectrum reconstruction under so-called supermoir\'e potentials. In Ref.~\citenum{kuiri_enhanced_2021} a strong reduction in Fermi-velocity was observed due to the proximity of two SDPs in a hBN/BLG/hBN supermoir\'e system. One therefore expects that specifics of the encapsulation such as the angle and the position of the hBN layers with respect to the encapsulated (multilayer) graphene can also be used to modify the bandgap both at the PDP and SDP, and that those effects may be strongly sensitive to gating. 

In this work we therefore investigate related mechanisms that will allow for the formation of gate-tunable flat bands based on precisely controlled stacking of hBN layers that encapsulate a multilayer graphene. We focus on the cases where the interference of moir\'e potentials on top and bottom graphene layers can be tuned, allowing for the formation of a gap at the SDP. Subsequent vertical gating of the system would then increase the gap at the PDP without significantly modifying the gap at the SPD, effectively flattening the moir\'e mini-band in between. As we will show, this mechanism is especially prominent in bilayer graphene and can indeed yield very flat dispersions, below 10~meV, using low gate voltages.

\section{Results and discussion}
\subsection{Theoretical model}
The alignment of graphene and hBN monolayers with 1.8$\%$ lattice mismatch results in a moir\'e pattern with a periodic length of $\lambda$ = 13.8~nm. As long as no twists are introduced in the heterostructure, the latter is the only length scale on which moir\'e effects will occur \cite{andelkovic_double_2020, kuiri_enhanced_2021}. The effects of the interfering moir\'e potentials on multilayer graphene can be easily understood from a simplified theoretical model. In the continuum limit, graphene on hBN can be described \cite{wallbank_generic_2013, wallbank_moire_2015, moon_electronic_2014, jung_origin_2015, jung_moire_2017} as a graphene layer upon which a moir\'e perturbation is applied, of the form
\begin{align}
    U_{eff}(\mathbf{r}) = \; &vG(u_0 f_1 + \tilde{u}_0 f_2)\sigma_0 + \zeta vG(u_3 f_1 + \tilde{u}_3 f_2) \sigma_3 \notag \\ 
    &+ \zeta v \left[ \mathbf{\hat{z}} \times \nabla (u_1 f_2 + \tilde{u}_1 f_1) \right. \notag\\
    &\left. + \nabla (u_2 f_2 + \tilde{u}_2 f_1)\right] \cdot\mathbf{\sigma}.
\label{eq:continuum_potential}
\end{align}
Here $\left(f_{1}(\mathbf{r}), f_{2}(\mathbf{r})\right)=\sum_{m}\left(1, i(-1)^{m-1}\right) \exp \left(i \mathbf{G}_{m} \cdot \mathbf{r}\right)$ are even and odd spatial functions along reciprocal lattice vectors $\mathbf{G}_{m} = \mathbf{b}_{gr,m} - \mathbf{b}_{hBN,m}$, where $\mathbf{b}_{gr,m}$ and $\mathbf{b}_{hBN,m}$ are the reciprocal lattice vectors of graphene and hBN. $\zeta$ is the valley index, $v$ the Fermi-velocity of monolayer graphene and $\sigma_i$ are the Pauli matrices with $\mathbf{\sigma} = [\zeta \sigma_1 , \sigma_2]$. The first term in equation~\eqref{eq:continuum_potential} is a spatially varying scalar potential, the second term is an induced mass term, and the third term represents the pseudo-vector potential due to the change in interlayer hoppings between graphene and hBN leading to a pseudo-magnetic field (PMF), with $u_i$ and $\tilde{u}_i$ being the spatially even and odd parts of each term. This effective model can be extended to the case of AB-stacked bilayer and ABC-stacked trilayer graphene as both have a 2-component low-energy representation $\mathcal{H}_{mult}$ in their low-energy sites, which are in both cases located in the top and bottom layer \cite{mccann_electronic_2013, zhang_band_2010}. The hBN layers, by good approximation, only apply a moiré potential as defined in equation \eqref{eq:continuum_potential} to the graphene layer that is directly adjacent, which for both BLG and TLG is the top and bottom layer: 
\begin{equation}
\tilde{\mathcal{H}} =   \mathcal{H}_{mult} + \left[ \begin{array}{cc}
    U_{b}(\mathbf{r})     &  0\\
    0     & U_{t}(\mathbf{r})
    \end{array}\right]. 
\label{eq:H_multilayer}
\end{equation}
Here $U_t(\mathbf{r})$ and $U_b(\mathbf{r})$ are the corresponding matrix elements for the low-energy sites from equation~\eqref{eq:continuum_potential} of the top (t) and bottom (b) layer. The effective potential for the multilayer graphene then becomes
\begin{align}
    U_{eff,mult}(\mathbf{r}) = &\; v_{mult}G(s_0 f_1 + \tilde{s}_0 f_2)\sigma_0 \nonumber \\
    &+ \zeta v_{mult}G(s_3 f_1 + \tilde{s}_3 f_2) \sigma_3.
\label{eq:continuum_potential_mult}
\end{align}
Here we neglect the term containing the pseudo-vector potential as it appears as a higher order correction for equation \eqref{eq:H_multilayer} which is discussed in section 1 of the supplementary material, but also to maintain the simplicity of the model. The scalar and the mass term for the multilayer graphene are now defined as a decomposition into $\sigma_0$ and $\sigma_3$ respectively in the basis of the low-energy sites positioned on top and bottom layers, such that applying different combinations of potentials on both layers can drastically change the components of equation~\eqref{eq:continuum_potential_mult}. We proceed to calculate the electronic bandstructure for bi- and trilayer graphene with an applied potential as defined in equation~\eqref{eq:continuum_potential_mult} using a real-space tight-binding approach in the moiré unit cell. In this manner, we report that a gap at the SDP does not appear in ABC trilayer graphene (TLG) when only considering the scalar and mass contributions as described in equation \eqref{eq:continuum_potential_mult}. However, using the same model, an effective gap at the SDP is obtainable in bilayer graphene (BLG). The $s_0$ term of equation~\eqref{eq:continuum_potential_mult} induces a gap at the SDP in either the electron or the hole side depending on the sign, $\tilde{s}_0$ induces an equally large gap in both electron and hole side simultaneously and so does $s_3$ together with flat band edges (reduced Fermi velocity) near the SDP, while $\tilde{s}_3$ does not open a gap at the SDP. Within this model, the condition to create an effective spectral gap is that the contributions of the components in \eqref{eq:continuum_potential_mult} work constructively both in amplitude and in sign such that the overlapping of bands is overcome. A more detailed discussion is provided in section 1 of the supplementary material. 

\begin{figure}[h]
    \centering
    \includegraphics{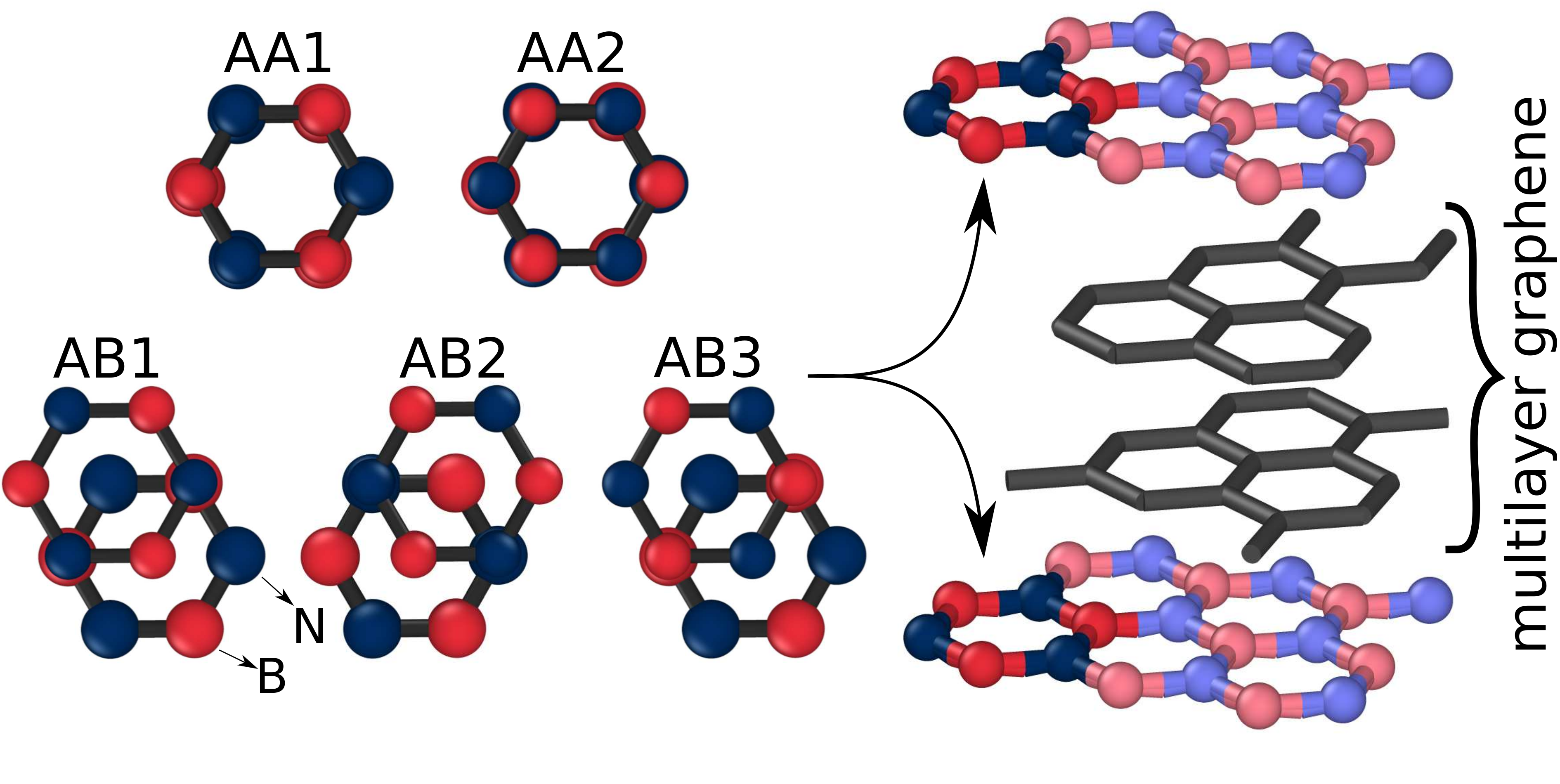}
    \caption{Nomenclature of stacking configurations of the hBN layers, for the encapsulation of multi-layer graphene.}
    \label{fig:stackings_hBN}
\end{figure}

\subsection{ABC-stacked trilayer graphene}

\begin{figure*}[t]
    \centering
    \includegraphics{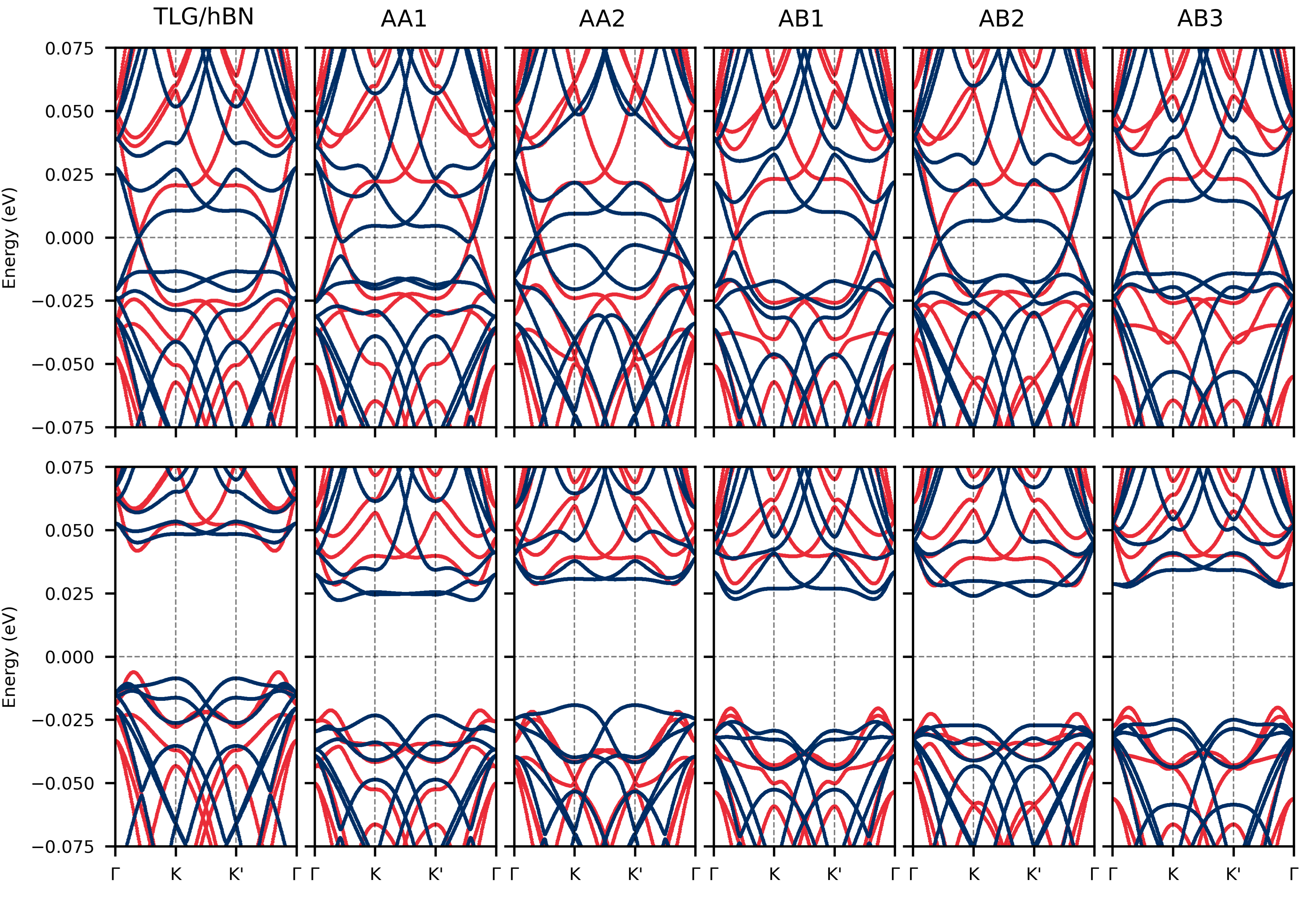}
    \caption{Electronic band-structures of ABC-stacked trilayer graphene, on a single hBN layer (TLG/hBN), and for the five different encapsulations with hBN (for hBN layers stacked as shown in figure~\ref{fig:stackings_hBN}), both for relaxed (blue) and unrelaxed (red) case. The bottom row shows the band-structure for the same systems under an applied vertical electric field of 100~mV/nm.}
    \label{fig:bands_hgggh}
\end{figure*}

Recent studies \cite{chen_evidence_2019, chen_signatures_2019} have shown that ABC-stacked TLG on hBN is a suitable platform for the formation of gate-tunable flatbands. While the ABA (Bernal) stacking is the preferred low energy configuration and most commonly occurring stacking for trilayer graphene, it is not consider in this work. In this case, the low energy bandstructure consists of a combination of linear and quadratic bands. In the presence of a perpendicular electric field, the overlap between these bands can be tuned but this does not a give rise to an electrically tuned gap at the PDP as it does in the ABC stacking configuration. ABC stacked trilayer graphene, while less abundant, is still readily obtainable and frequently used in experiments. In light of this, we perform real-space tight-binding calculations for the whole moir\'e supercell of the encapsulated ABC trilayer graphene in which now the hBN layers are added in an atomistic way. We further include relaxation effects by performing semi-classical molecular dynamics simulations. The absence of gaps in our simplified model \eqref{eq:continuum_potential_mult} and the lack of inclusion of the pseudo-vector potential term, indicates the importance of the PMF in the formation of gaps. Relaxation effects on the other hand are known to induce a large PMF  due to intralayer hopping modulation \cite{jung_origin_2015, jung_moire_2017, long_accurate_2021}, thus becoming a crucial ingredient in this calculation. In what follows, we select five highly symmetric stacking configurations of the hBN layers as shown in figure~\ref{fig:stackings_hBN}, to preserve the highest degree of symmetry, and which are all aligned as we are interested only in band reconstruction at a singular energy level. 

\begin{figure*}[t]
    \centering
    \includegraphics{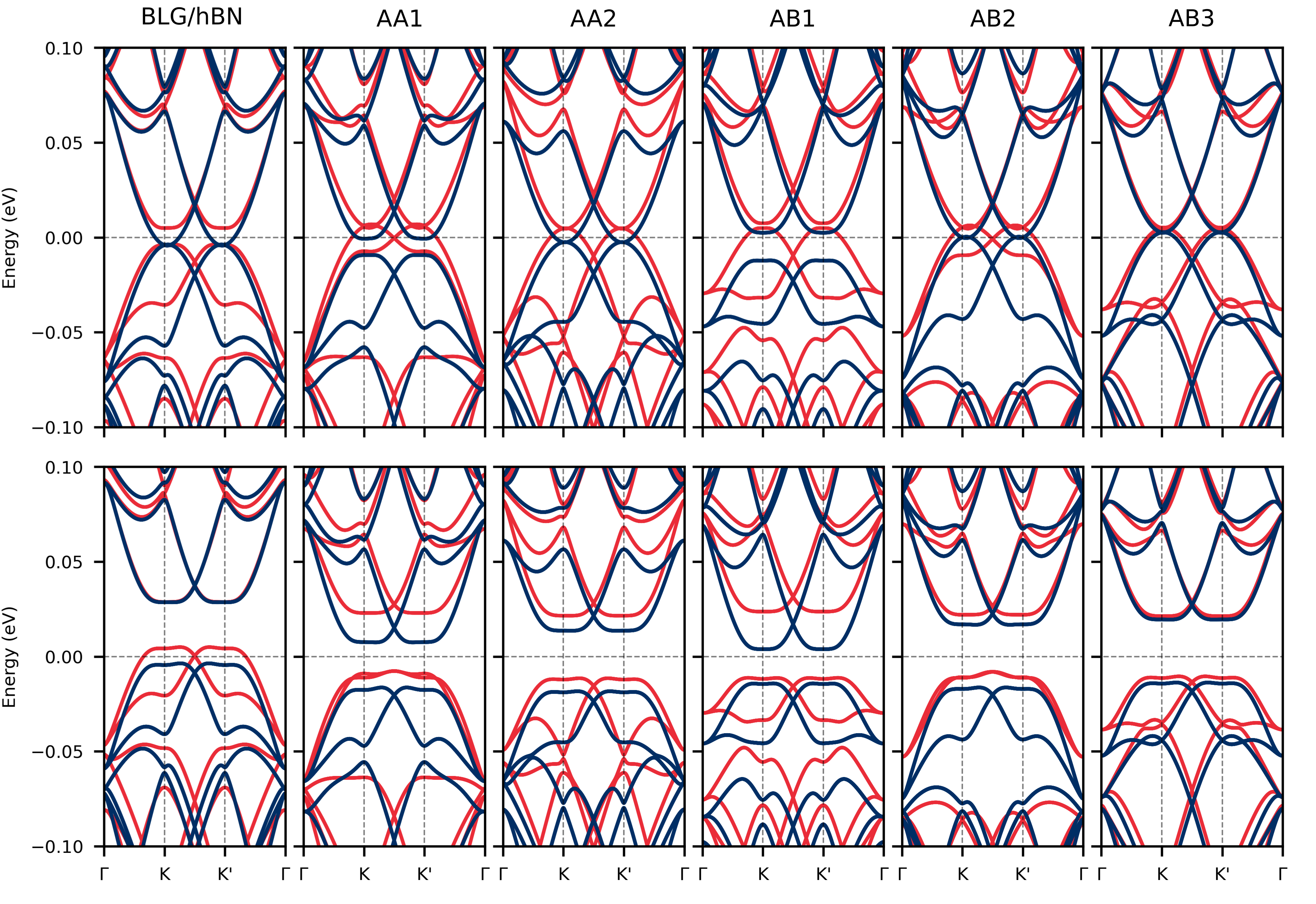}
    \caption{Electronic band-structures of AB-stacked bilayer graphene, on a single hBN layer (BLG/hBN), and for the five different encapsulations with hBN (for hBN layers stacked as shown in figure~\ref{fig:stackings_hBN}), both for relaxed (blue) and unrelaxed (red) case. The bottom row shows the band-structure for the same systems under an applied vertical electric field of 100~mV/nm.}
    \label{fig:bands_hggh}
\end{figure*}

The resulting electronic band-structures for the encapsulated ABC-stacked TLG configurations are shown in figure~\ref{fig:bands_hgggh}. The Dirac cone is now positioned between the original K-point and the $\Gamma$-point due to a Lifshitz transition near the Dirac point coming from the $\gamma_{2}$ hopping between the top and the bottom graphene layer, and is a typical feature seen in ABC TLG \cite{zhang_band_2010}. No effective spectral gap is observed for the unrelaxed systems besides the band anti-crossing near the $\Gamma$-point in some cases and a tiny gap at the PDP when inversion symmetry is broken. The emergence of the band anti-crossing is presumably due to the PMF-term in equation \ref{eq:continuum_potential} as no such gap was observed in our analysis based on the simplified model, shown in the supplementary material section 1. It is only after relaxation effects are taken into consideration that gaps emerge at the SDP. This supports our preceding intuition that the gaps are enhanced by the PMF, as relaxation induces a large PMF in each graphene layer. Previous work has shown the presence of a gate-tunable isolated flat band in TLG on monolayer hBN (TLG/hBN) and the emergence of highly correlated states  both in experiment and theory \cite{chen_evidence_2019, chen_signatures_2019}. The band in question is the first valence band, which in our model reaches bandwidth of 17~meV (38~meV) for the relaxed (unrelaxed) case, for the same gating value that induces a potential difference of 10~meV between the top and the bottom graphene layer. There is however some slight overlap of the valence band with lower-lying bands, i.e. it is not fully isolated. For the other stacking configurations with relaxation included we observed similar flattening of the conduction band, the smallest bandwidth being 11~meV for AA1 stacking at 100~mV/nm, 11~meV for AB3 at 150~mV/nm, and 5~meV for TLG/hBN at 150~mV/nm gating field. We note that the magnitude of the gaps is dependent on the strength of the coupling to the hBN layers. In section 2 of the supplementary material we show results with a different parameter set for the coupling between graphene and hBN. We find that the gaps at the PDP and SPD are smaller and the bandwidth is larger when the coupling is weaker. In order to have a precise quantitative prediction of the induced gaps, tight-binding parameters based on accurate ab-initio simulations of the graphene/hBN interface are needed.

\begin{figure*}[t]
    \centering
    \includegraphics{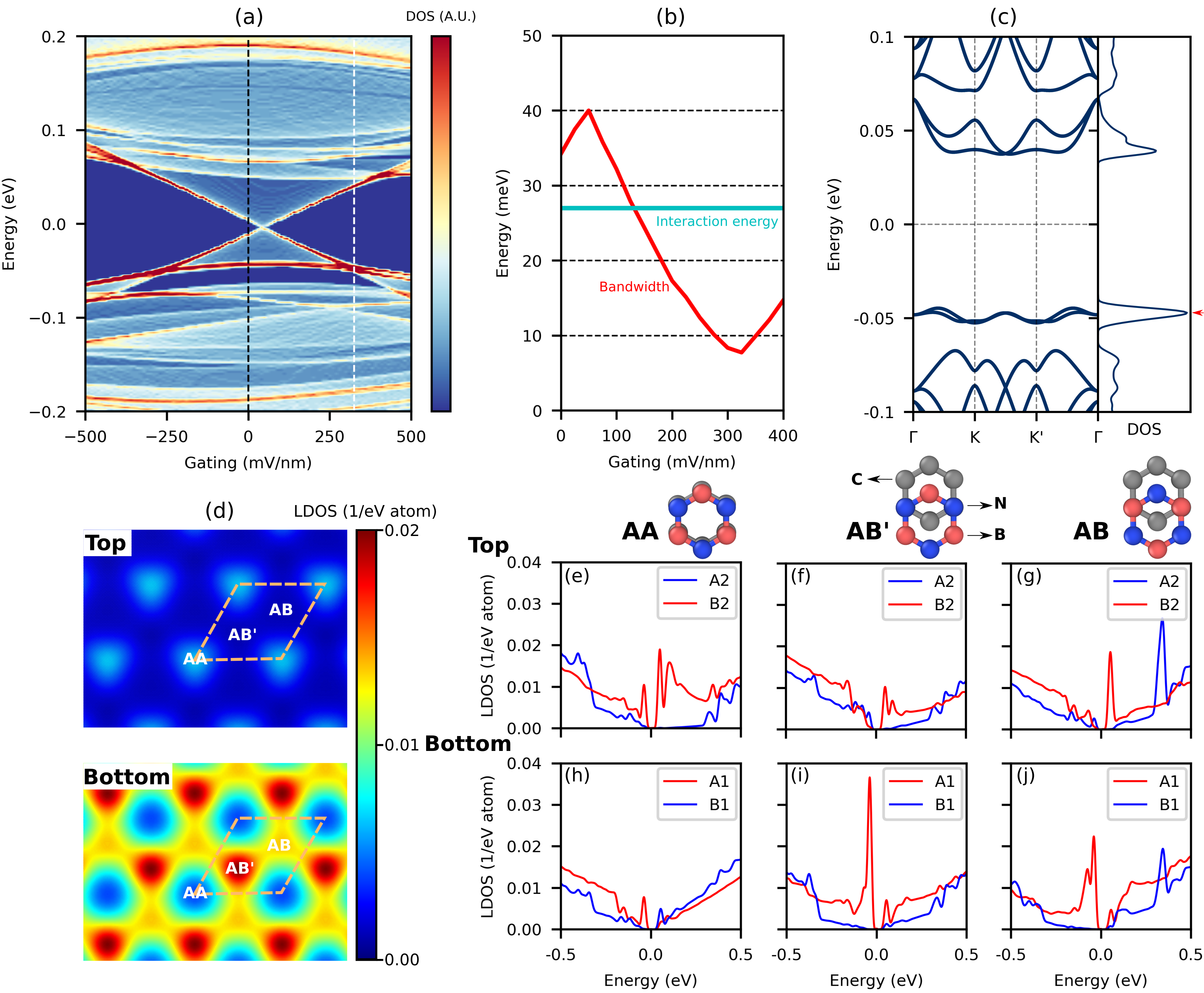}
    \caption{(a) Average DOS of the relaxed AB1-encapsulated BLG taken along the $\Gamma$-K-K'-$\Gamma$ path as a function of applied gating. The gating values of 0 and 325~mV/nm are indicated by a black and white dashed line respectively. (b) Coulomb interaction energy (blue) of the charge carriers and bandwidth (red) of the first valence band of the relaxed AB1 system, where a minimum bandwidth of 7~meV is reached at 325~mV/nm gating. (c) Band structure of the relaxed AB1 structure for applied electric field of 325~mV/nm together with the DOS (2~meV broadening). The red arrow indicates the first valence band. (d) The LDOS on the low-energy sites within the graphene layers of the AB1 structure, under 325~mV/nm gating, and at -0.05 eV below the Dirac point, given in number of states per eV per atom. (e-j) The LDOS in the center of stacking regions AA, AB' and AB of (d), for the top and bottom layer. AA stacking corresponds to carbon being on top of both boron and nitrogen, AB stacking corresponds to carbon being on top of only boron, and in AB' stacking carbon is positioned on top of only nitrogen. Here A1 and B2 are the low-energy sites (red) and B1 and A2 the high-energy (dimer) sites (blue). Note that a smaller broadening (5~meV) was used for plots (e)-(j) than in panel (d) (20~meV), which causes difference in peak LDOS values.}
    \label{fig:data_AB1}
\end{figure*}

\subsection{AB-stacked bilayer graphene}
We next perform similar atomistic tight-binding calculations for the hBN-encapsulated configurations of AB-stacked bilayer graphene. The resulting band-structures are shown in figure~\ref{fig:bands_hggh}. For one configuration, the AB1 stacking of hBN, a large effective gap appears at the SDP that is robust under relaxation. For this configuration the effective scalar moir\'e potentials on top and bottom graphene layer are identical, with opposite sign for the intralayer mass terms $u_3$ and $\tilde{u}_3$, which maximizes the amplitude of the $s_3$ moir\'e term in the 2-component model for BLG and leading to strongly reduced Fermi-velocity at the SDP. The gap also remains unchanged when gating is applied and the gap at the PDP increases, which is apparent from figure~\ref{fig:data_AB1}(a) where the DOS is zero at a nearly constant energy between gating values of $\approx \pm$ 350~mV/nm. A similar gap is seen for AB2, but closes after relaxation is included. In general we also find that a small gap at the PDP appears whenever the inversion symmetry is broken, which is the case for BLG/hBN, AA1 and AB1 configurations. After the structures are relaxed, these gaps get enhanced even further, except for BLG/hBN in which the gap becomes vanishingly small, which is in agreement with experimental observations \cite{kim_accurate_2018}. For these systems the application of a perpendicular gate potential will first close the gap at the PDP as it counteracts the average non-zero mass term, after which the gap is opened up again - as seen in figure~\ref{fig:data_AB1}(a). The broken inversion symmetry and gap at the PDP in the BLG/hBN, AA1 and AB1 configurations leads to layer polarization which could further indicate the presence of unconventional ferroelectricity as was recently observed experimentally  \cite{zheng_unconventional_2020} and  described theoretically \cite{zhu_electric_2022} in twisted hBN/BLG/hBN systems. For completeness, in section 4 of the supplementary material, we provide the DOS of each configuration for each sublattice, showing the layer polarization in BLG/hBN, AA1 and AB1.

We next turn the attention to the gap at the SDP in the AB1 BLG configuration, that will squeeze the first valence band in width in the presence of gating. As plotted in figure~\ref{fig:data_AB1}(b), the bandwidth decreases down to 7~meV for an applied gating field of 325~mV/nm, which corresponds to a potential difference of 325~mV across the thickness of the sample ($\approx$1~nm). Such electric field values across a thin 2D heterostructure lay well within reach of experimental capabilities. Most recently, Refs.~\citenum{domaretskiy_quenching_2022} and \citenum{weintrub_generating_2022} reported realizations of an electric field up to 3.5~V/nm across a $WSe_2$ bilayer using ionic liquid gates, one order of magnitude higher than required in this work. In figure~\ref{fig:data_AB1}(c) the band structure for the relaxed AB1 BLG configuration is presented with an applied electric field of 325~mV/nm, clearly showing the well-isolated first valence band and its ultra-flat dispersion. The effective gap to the second valence band is about 15~meV. The narrow bandwidth comes with a strong localization of the electronic wavefunction. The local density of states (LDOS) at -0.05 eV for the relaxed AB1 system with gating of 325~mV/nm is shown in figure~\ref{fig:data_AB1}(d), together with the LDOS for the energy range of -0.5 eV to 0.5 eV at the different local stacking arrangements in the top and bottom layer in panels (e)-(j) of the same figure. The wavefunction is strongly localized at the low-energy sites in the AB' regions in the bottom layer. This is due to the gating, which induces a negative potential on the bottom and positive potential on the top layer. The Coulomb interaction energy can be estimated by taking the distance between charge carriers as the distance between the AB' regions, which is nothing more than the periodic length $\lambda = 13.8$~nm of the moir\'e pattern. We find then that $U = \frac{e^{2}}{4 \pi \epsilon_0 \epsilon L_M} \approx 27$~meV, where $\epsilon = 4$ is the dielectric constant of hBN and $L_M \approx \lambda$ the inter-particle distance. In figure~\ref{fig:data_AB1}(c) we compare this energy to the bandwidth, serving as an estimate for the kinetic energy of the charge carriers. One sees that upon gating $U>E_{kinetic}$, so that we enter a regime where the interaction energy dominates, which is expected to lead to highly-correlated electron-electron states in the first valence band. This is similar to magic-angle twisted bilayer graphene, where the localization is in the AA stacked regions with a moir\'e length of 13~nm, as well as TLG/hBN where the moir\'e length is identical.

\subsection{Structural relaxation}
\begin{figure}[h]
    \centering
    \includegraphics{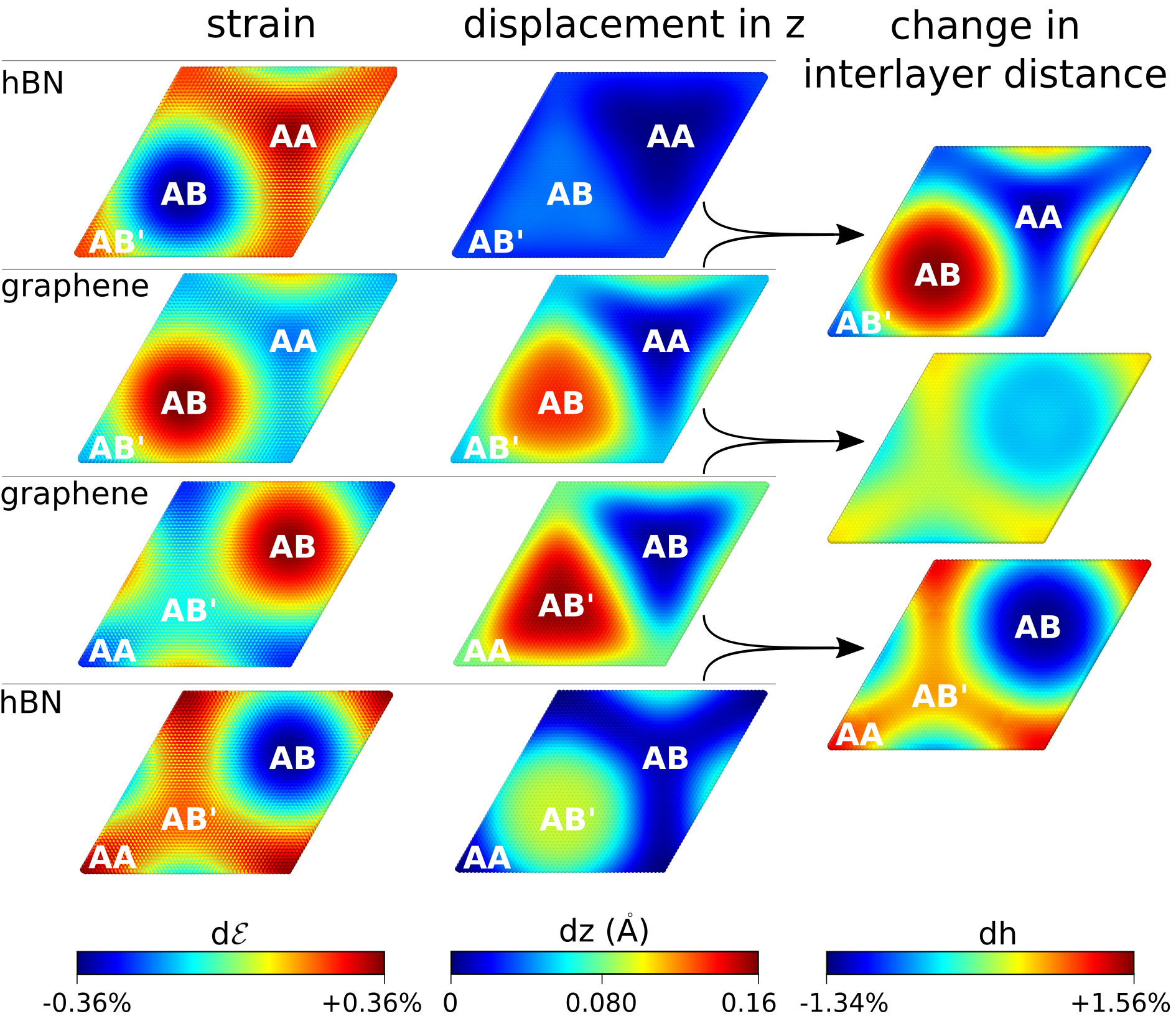}
    \caption{Local strain (left panels) and the out-of-plane displacement (central panels) upon relaxation of the AA1-encapsulated BLG structure. The displacement values are taken in reference to the lowest $z$-coordinate of each layer when unrelaxed. The right panels show the percentage-wise change in interlayer distances in reference to the original interlayer distances.}
    \label{fig:strain}
\end{figure}

To better understand the effects of structural relaxation on the electronic properties we provide a qualitative analysis of the modifications due to the relaxation of the AA1-encapsulated bilayer graphene, plotted in figure~\ref{fig:strain}. It is important to note that in order to couple the system to a flat substrate as is often the case in electronic devices and in order to suppress out-of-plane buckling, the top hBN layer is coupled to a 1D 12/6-Lennard-Jones potential $V_{LJ}(z)$, with parameters $\epsilon = 0.05$ and $\sigma = 3.5$. The relaxation strain forms a trigonal pattern, where in the regions with AB stacking, the graphene layers will stretch the most while the hBN layers will compress the most, decreasing the lattice mismatch and effectively increasing the AB region in size. Within the surrounding AA and AB' stacking regions the opposite emerges, where now the graphene layers compress and the hBN layers expand, the compression (expansion) is the largest in the AA region within the graphene (hBN) layers and extends along lines connecting AA and AB' regions. This again means that within the AA- and to a lesser extent AB'-regions the lattice mismatch is increased and the AA- and AB'- regions decrease in size. The AB stacking is the most energetically favorable while the AA stacking is the least favorable one, so that the expansion and compression of the AB and AA regions respectively minimizes the energy \cite{argentero_unraveling_2017, woods_commensurateincommensurate_2014, jung_origin_2015}. The strain within the graphene layer is in good agreement with Ref.~\citenum{argentero_unraveling_2017} while the strain within the hBN layers here ranges from -0.39\% to +0.36\%, which is larger than in Ref.~\citenum{argentero_unraveling_2017} where it ranges from -0.15\% to +0.20\%. This discrepancy can be largely attributed to the suppression of out-of-plane buckling due to the Lennard-Jones interaction with the top hBN layer. The strain of adjacent layers also seems to have little-to-no influence on the strain in other layers in the structure. In the supplementary material section 6 we show that the resulting modulation of the hoppings gives rise to a PMF on the order of 9~T. The out-of-plane displacement of individual layers in this case reaches values of up to 0.082~\AA, with the effect being much stronger in the graphene layers than in the hBN layers. Moreover, the displacement seems to be always localized in the same place for each layer. In our case the buckling is determined by the stacking of the bottom graphene and hBN layers, as the bottom hBN layer does not contain a Lennard-Jones potential. In the bottom graphene and hBN layer, the AB-stacked regions buckle with the concave side in the graphene layer and the AB'-stacking region with the concave side in the hBN layer, also in agreement with Ref.~\cite{argentero_unraveling_2017}. The hBN-graphene interlayer distance varies on the order of $\pm1.5\%$ of the original interlayer distance, which modulates the hoppings by approximately 0.03$\gamma_1$ or 0.01~eV, making it a rather weak effect. Especially the observed variation in interlayer distance between the graphene layers is insignificantly small.

\section{Conclusion}
In conclusion, the hBN-encapsulated AB-stacked bilayer graphene was shown to be an ideal platform for the generation of low-dispersive bands and strong localization of electronic states, where stacking of hBN layers in combination with gating leads to a new mechanism for flattening the bands and tailoring the strongly-correlated states. By individually tuning the gap at the primary and the secondary Dirac point (by gating and by stacking of hBN layers, respectively), we were able to generate flat bands with a bandwidth below 10~meV, significantly lower than the estimated Coulomb interaction energy - hence facilitating the emergence of highly-correlated electron states. The achieved bandwidth is similar to that of twisted BLG \cite{cao_unconventional_2018, cao_correlated_2018} ($\approx$ 5-10~meV) and TLG/hBN \cite{chen_signatures_2019, chen_evidence_2019} ($\approx$ 11.7~meV) where strongly-correlated phases have readily been observed and where the localization of charge carriers was on a similar length scale as was found here (so that the on-site Coulomb energy is similar as well). Encapsulating graphene with hBN layers is readily used to drastically improve the electronic properties of graphene, but our findings reveal that stacking of hBN layers is a relevant and potent ingredient to achieve high quality devices with gate-tunable strongly-correlated electronic states.

\section{Methods}
\subsection{Tight-binding model}
The system is described by a tight-binding model, where a Bloch wavefunction of the form
\begin{equation}
\left|\Psi_{\mathbf{k}}\right\rangle= \frac{1}{\sqrt{N}}\sum_{i}^{N} e^{i \mathbf{k} \mathbf{R}_{i}}\left|\phi_{i}\right\rangle,
\end{equation}
is used with wavefunction $\left|\phi_{i}\right\rangle$ of atomic site i at position $\mathbf{R}_{i}$. The most general form of the Hamiltonian is then given by
\begin{equation}
    H = -\sum_{i,j} t(\mathbf{R}_i - \mathbf{R}_j) |\phi_{i}\rangle \langle \phi_{j} | + \sum_{i}V(\mathbf{R}_{i}) | \phi_{i} \rangle \langle \phi_{i} |. \label{eq:general_H}
\end{equation}
The first term in equation~\eqref{eq:general_H} defines the hopping between any sites i and j with a hopping strength t. The second term represents the potential due to the on-site potentials ($V_{C} = 0$ for carbon, $V_{B} = 3.34$~eV for boron and $V_{N} = -1.40$~eV for nitrogen). The matrix representation in the basis of the wavefunctions $|\phi_{i} \rangle$ is then 
$H_{i j} = \left\langle\phi_{i}|H| \phi_{j}\right\rangle$.

We make use of the Slater-Koster (SK) type of functions \cite{slater_simplified_1954}, which, depending on the type of orbitals and their relative positions, give a good estimate of the hopping value. In this work we closely follow the approach and parameters as described in Ref.~\citenum{moon_electronic_2014} , leading to
\begin{equation}
-t(\mathbf{R})=V_{p p \pi}\left[1-\left(\frac{\mathbf{R} \cdot \mathbf{e}_{z}}{R}\right)^{2}\right]+V_{p p \sigma}\left(\frac{\mathbf{R} \cdot \mathbf{e}_{z}}{R}\right)^{2},
\label{eq:hopping_formula}
\end{equation}
where
\begin{align}
    &V_{p p \pi}=V_{p p \pi}^{0} \exp \left(-\frac{R-a_{0}}{r_{0}}\right), \label{eq:exp_decay_pi} \\
    &V_{p p \sigma}=V_{p p \sigma}^{0} \exp \left(-\frac{R-d_{0}}{r_{0}}\right). \label{eq:exp_decay_sigma}
\end{align}
Here $r_{0}$ = 0.026~nm is the decay length, $a_{0}$ = 0.142~nm the interatomic distance in graphene, $d_{0}$ = 0.335~nm the interlayer distance between graphene layers, $V_{p p \sigma}^{0}$ = 0.48~eV, and $V_{p p \pi}^{0}$ = -2.7~eV. The exponential decay is set up in such a way that nnn hopping equals 0.1$V^{0}_{pp \pi}$, given that in most literature $\gamma_{nnn} \approx 0.1 \gamma_{0}$ for graphene. This specific adjustment was respected to further determine all the hoppings within the bilayer graphene systems for both unrelaxed and relaxed systems.

For ABC TLG another set of parameters was used due to the presence of next-nearest layer hoppings. From Ref.~\citenum{menezes_ab_2014} we obtained $\gamma_0$ = -2.577~eV, $\gamma_1$ = 0.348~eV, $\gamma_2$ = -0.024~eV, $\gamma_3$ = 0.290~eV, $\gamma_4$ = 0.196~eV and $\gamma_{nnn}$ = -0.258~eV. For the hopping between neighbouring graphene and hBN layers, we rely again on the use of the SK functions to estimate the hopping parameters and adjust $V_{pp\sigma}^0$ and $V_{pp\pi}^0$ in \eqref{eq:exp_decay_pi} and \eqref{eq:exp_decay_sigma} accordingly.
For the graphene in the relaxed systems we then use:
\begin{equation}
    t_{i}(\mathbf{R}) = \gamma_{i} \frac{t_{SK}(\mathbf{R})}{t_{SK}(\mathbf{R}_{0})}, 
    \label{eq:hopping_tlg}
\end{equation}
where $t_{SK}(\mathbf{R})$ is the hopping as calculated in equation \eqref{eq:hopping_formula}, and $\mathbf{R}_{0}$ is the atomic separation for the unrelaxed system.

A symmetric top-back gating is applied by introducing a potential field along the $z$ direction $V(z) = V_{0}z$, where in all the atomic sites within the same layer the same potential was applied.

In order to set up and solve the tight-binding model we made use of Pybinding \cite{pybinding}, which uses an efficient routine to setup and solve the band-structure, and relies on the KPM method \cite{weise_kernel_2006} in order to obtain the electronic DOS. In this work a Gaussian broadening of 5~meV width was used in the KPM formalism (unless mentioned otherwise). A cutoff range for hopping-pairs to form was set to 1.75 times the graphene inter-atomic distance $a_{cc}$ for intralayer hoppings, and 1.40 times the interlayer distance between graphene layers $C_{0}$ for interlayer hoppings. Next-nearest neighbours (nnn) in the graphene layers were included to improve the accuracy of our calculations and possibly capture any effects, such as electron-hole asymmetry, that might occur. It is known that the inclusion of nnn results in an energy shift of the Dirac point equivalent to three times the next-nearest-neighbor hopping $\gamma_{nnn}$, therefore a shift in energy of the same amount was applied such that the Dirac point sits at the Fermi level, which is 0.774~eV for TLG and 0.81~eV for BLG.

\subsection{\label{sec:level2}Relaxation}
The relaxation was performed through a classical molecular dynamics simulation as implemented in LAMMPS \cite{plimpton_fast_1995, plimpton_computational_2012}, for all the layers simultaneously. For intra- and interlayer carbon-carbon interactions we used the bond-order Brenner potentials \cite{brenner_second-generation_2002} and Kolmogorov-Crespi potentials \cite{kolmogorov_smoothest_2000} respectively. For the boron-nitrogen intralayer interactions we used Tersoff potentials \cite{tersoff_new_1988}, and the Morse potential for the interaction between graphene and hBN layers \cite{argentero_unraveling_2017}. For the Morse potential we used as equilibrium distances 3.54~\AA \; for B-C and 3.52~\AA \; for N-C. In order to mimic the effects of a substrate, a 12/6-Lennard-Jones potential $V_{LJ}(z)$, with parameters $\epsilon = 0.05$ and $\sigma = 3.5$, is used to couple the system to a flat surface. This suppresses the out-of-plane buckling that otherwise commonly occurs in these systems. A single moir\'e supercell with periodic boundary conditions is then simulated at zero temperature starting from the unrelaxed configuration. The total energy of the system is minimized until the forces reach a value below $10^{-6}$~eV/\AA.

\section*{Conflicts of interest}
There are no conflicts to declare.

\begin{acknowledgments}
The authors would like to acknowledge the contribution of Mi\v{s}a Andelkovi\'c, providing invaluable support leading up to this work.  The work was supported by the Research Foundation - Flanders (FWO-Vl), project number G0A5921N. The computational resources were provided by the HPC core facility CalcUA of the Universiteit Antwerpen, and VSC (Flemish Supercomputer Center).
\end{acknowledgments}

\bibliography{My_Library_modified}% Produces the bibliography via BibTeX.

\end{document}

\title[Supporting information]{Supporting information for: "Strong gate-tunability of flat bands in bilayer graphene due to moir\'e encapsulation between hBN monolayers"}

\author{Robin Smeyers, Milorad V. Milo\v{s}evi\'c, Lucian Covaci}
\eads{\mailto{robin.smeyers@uantwerpen.be}, \mailto{milorad.milosevic@uantwerpen.be}, \mailto{lucian.covaci@uantwerpen.be}}
\address{Department of Physics and NANOlab Center of Excellence, University of Antwerp, Groenenborgerlaan 171, 2020 Antwerp, Belgium}
\vspace{10pt}
\begin{indented}
\item \date{\today}
\end{indented}

%\noindent{\it Keywords}:
% \submitto{\TDM}
%\maketitle

\section{Analysis of moir\'e terms}
\label{sec:moire_analysis}
In this section we provide a numerical analysis of the different terms present in the effective moir\'e potential. The moir\'e potential that emerges from placing a single hBN layer onto a single graphene layer is given in general by \cite{wallbank_generic_2013}:
\begin{align}
    U_{eff}(\mathbf{r}) = \; &vG(u_0 f_1 + \tilde{u}_0 f_2)\sigma_0 + \zeta vG(u_3 f_1 + \tilde{u}_3 f_2) \sigma_3 \notag \\ 
    &+ \zeta v \left[ \mathbf{\hat{z}} \times \nabla (u_1 f_2 + \tilde{u}_1 f_1) + \nabla (u_2 f_2 + \tilde{u}_2 f_1)\right] \cdot\mathbf{\sigma}.\notag\\
\label{eq:continuum_potential}
\end{align}
where 
\begin{equation}
\left(f_{1}(\mathbf{r}), f_{2}(\mathbf{r})\right)=\sum_{m}\left(1, i(-1)^{m-1}\right) \exp \left(i \mathbf{G}_{m} \cdot \mathbf{r}\right)
\end{equation}
are the even and odd periodic functions in which the potential can be decomposed and $\zeta$ is the valley index. The first term is the scalar potential, the second term describes the mass term indicating the potential difference between the A and B sublattice, and the third term represents the pseudo-vector potential due to the local modification of the interlayer hopping values. $u_i$ and $\tilde{u}_i$ are the even and odd part of the corresponding Pauli decomposition in matrices $\sigma_i$ and where $\mathbf{\sigma} = \left[ \zeta \sigma_1, \sigma_2 \right]$.

Since we are mainly interested in the electronic properties at low energy we can now switch to the 2-component model for AB-stacked BLG and ABC-stacked TLG in the basis of its corresponding low-energy sites. The low-energy sites are in both cases located on top and bottom layer. The transformation to the low energy model is almost identical for both systems so that here we will only derive the low-energy model for BLG \cite{mccann_electronic_2013} and the one for TLG can be derived in an analogous way as described in Ref.~\cite{koshino_trigonal_2009}. We can split up the original Hamiltonian for BLG into components $\theta$=$(\psi_{A1}, \psi_{B2})$ and $\chi$=$(\psi_{A2}, \psi_{B1})$ for the low- and high-energy sites of the system:

\begin{equation}
\left(
\begin{bmatrix}
h_{\theta} & D^{\dagger} \\
D & h_{\chi}
\end{bmatrix}
\; + \; U_{t+b} \right)
\begin{bmatrix}
\theta \\
\chi
\end{bmatrix}
= E 
\begin{bmatrix}
\theta \\
\chi
\end{bmatrix},
\label{eq:abblg_low_energy_eq}
\end{equation}
 where $U_{t+b}$ is a 4$\times$4 matrix representing the applied moiré potential on top and bottom layer in the same basis (A1, B2, A2, B1):

\begin{equation}
    U_{t+b} = 
    \begin{bmatrix}
    V_b(\mathbf{r}) + M_b(\mathbf{r}) & 0 & 0 & A_{1,2}(\mathbf{r}) \\\\
    0 & V_t(\mathbf{r}) - M_t(\mathbf{r}) & A_{4,3}(\mathbf{r}) & 0 \\\\
    0 & A_{3,4}(\mathbf{r}) & V_t(\mathbf{r}) + M_t(\mathbf{r}) & 0 \\\\
    A_{2,1}(\mathbf{r}) & 0 & 0 & V_b(\mathbf{r}) - M_b(\mathbf{r})
    \end{bmatrix}
    =
    \begin{bmatrix}
    V_{1,1} & V_{2,1}^{\dagger} \\
    V_{2,1} & V_{2,2}
    \end{bmatrix}
\end{equation}
Here $V_i(\mathbf{r})$, $M_i(\mathbf{r})$ and $A_{k,l}(\mathbf{r})$ are the scalar, mass and vector potentials from equation \ref{eq:continuum_potential}. The moiré potential splits up into a block-form where the scalar and mass terms appear together on the diagonal and the vector potential terms appear on the off-diagonal blocks. We can further reduce equation \ref{eq:abblg_low_energy_eq} to an equation in the low-energy eigenstates $\theta$ and by expanding until first order around low E:

\begin{align}
    \left(h_{\theta} + V_{1,1} - (D + V_{2,1}) ^{\dagger} (h_{\chi} + V_{2,2}) ^{-1} (D + V_{2,1})\right) \theta =& E \tilde{Q}\theta \label{eq:moire_perturbation1} \\
    \left(\tilde{h}_{\theta} - \tilde{D} ^{\dagger} \tilde{h}_{\chi} ^{-1} \tilde{D}\right) \theta =& E \tilde{Q}\theta \label{eq:moire_perturbation2}
\end{align}
Where:
\begin{align}
    \tilde{Q} &= 1 + \tilde{D}^{\dagger} \tilde{h}_{\chi}^{-2}\tilde{D} \\
    \tilde{h}_{\theta} &= h_{\theta} + V_{1,1} \\
    \tilde{h}_{\chi} &= h_{\chi} + V_{2,2} \\
    \tilde{D} &= D + A_{2,1}
\end{align}
From equation \eqref{eq:moire_perturbation2} we can identify that $V_{1,1}$, containing the added on-site potential on the low-energy sites is the main contribution in the moir\'e potential and that the vector potentials and on-site potentials on the high-energy sites are a higher order correction. In light of this we further discuss the effects of the $V_{1,1}$-term which can now again be decomposed into a scalar and a mass term, with the latter being the potential difference between the low-energy sites. We get in the most general form:

\begin{align}
    \tilde{V}(\mathbf{r}) =& \; V_0 +  S_V(\mathbf{r}), \label{eq:general_moire_scalar} \\
    \tilde{M}(\mathbf{r}) =& \; M_0 +  S_M(\mathbf{r}),
    \label{eq:general_moire_mass}
\end{align}
where $V_0$ and $M_0$ are a constant for the scalar and mass term and $S_V(\mathbf{r})$ and $S_M(\mathbf{r})$ are the spatially varying parts with period of the original moir\'e potential as no twist is applied between the layers. 
$V_0$ will cause a general shift of the spectrum, while $M_0$ will generate a bandgap at the primary Dirac point (PDP) due to breaking of sublattice symmetry. The spatially varying parts of the 2-component model can be split up into even and odd components just like in equation \eqref{eq:continuum_potential}:
\begin{align}
    S_V(\mathbf{r}) =& v_{mult}G(s_0 f_1 + \tilde{s}_0 f_2) \sigma_0, \label{eq:scalar_spatial_moire} \\ 
    S_M(\mathbf{r}) =& v_{mult}G(s_3 f_1 + \tilde{s}_3 f_2) \sigma_3.
\label{eq:mass_spatial_moire}
\end{align}
Here we provide a numerical analysis of the different components in equation~\eqref{eq:mass_spatial_moire} and \eqref{eq:scalar_spatial_moire} by applying the appropriate spatially varying potential onto a pristine BLG in the moir\'e unit cell. Our main point of interest is the behaviour of the gap at the secondary Dirac point (SDP) and how it is modified. It is important to note that for aligned monolayer graphene on hBN an analytical expression for the gap at the SPD has previously been derived \cite{wallbank_moire_2015}.

\begin{figure}
    \centering
    \includegraphics{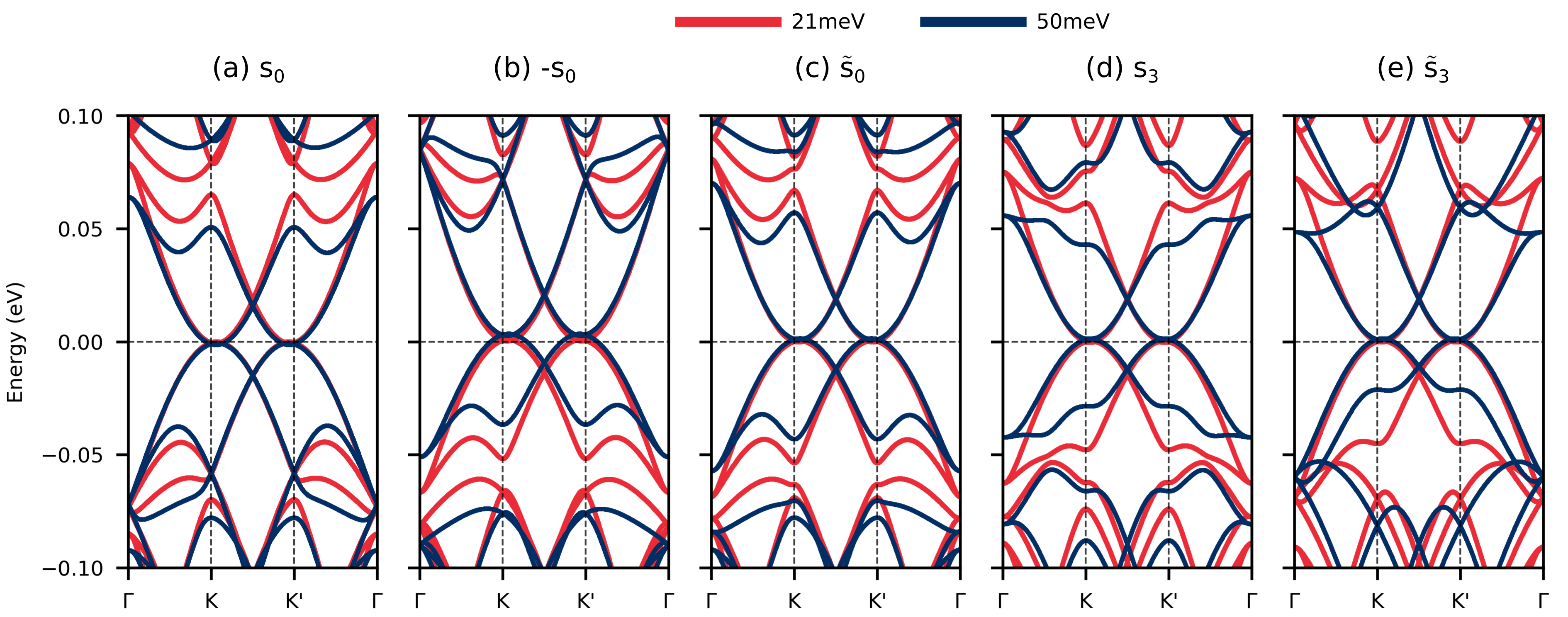}
    \caption{Band structure of an AB BLG supercell upon which a periodic potential has been applied that corresponds to the different moir\'e perturbation terms of the 2-component model. The potential is only applied on the low-energy sites with an amplitude of 21~meV (red) and 50~meV (blue). (a) The applied potential is strictly even and of the form $\sum_i V_0 \cos (\mathbf{G}_i.\mathbf{r})$ and identical on both layers. (b) The same as (a) but with a negative sign. (c) The applied potential is of the form $\sum_i V_0 \sin (\mathbf{G}_i.\mathbf{r})$ and with the same sign between both layers. (d) Potential is of the form $\sum_i V_0 \cos (\mathbf{G}_i.\mathbf{r})$ that is opposite on both layers. (e) The potential is of the form $\sum_i V_0 \sin (\mathbf{G}_i.\mathbf{r})$ and with opposite sign between both layers.}
    \label{fig:moire_bands_blg}
\end{figure}

In figure~\ref{fig:moire_bands_blg} the bandstructure of AB-stacked BLG in the moir\'e unit cell of unrotated graphene/hBN is given upon which a periodic potential is applied that corresponds with the different contributions in (\ref{eq:scalar_spatial_moire}) and (\ref{eq:mass_spatial_moire}). The $s_0$ term is the even part of the scalar potential and shows a gap only on the electron side or the hole side depending on the sign of the potential, suggesting that most electron-hole asymmetry comes from this term. The $\tilde{s}_0$ is the odd part of the scalar potential and shows the formation of a gap at the SDP that is equal in magnitude for hole and electron side. The even part of the mass term is given by $s_3$ and again shows the formation of a gap at the SDP that is largely preserving the electron-hole symmetry and inducing flatter band edges. Lastly we have $\tilde{s}_3$ which is the odd part of the mass term and shows no induced gap at the SDP. The mass term in general appears to induce some strong modification of the bands near the K-point in the first valence band for $\tilde{s}_3$ and in both valence and conduction band for $s_3$. The only term that is sign-dependent is $s_0$, which is easy to understand since the odd functions are by definition equal in magnitude and shape for positive and negative regions and the mass term is also only dependent on the absolute difference in potentials which a switch of sign will not change.

\begin{figure}
    \centering
    \includegraphics{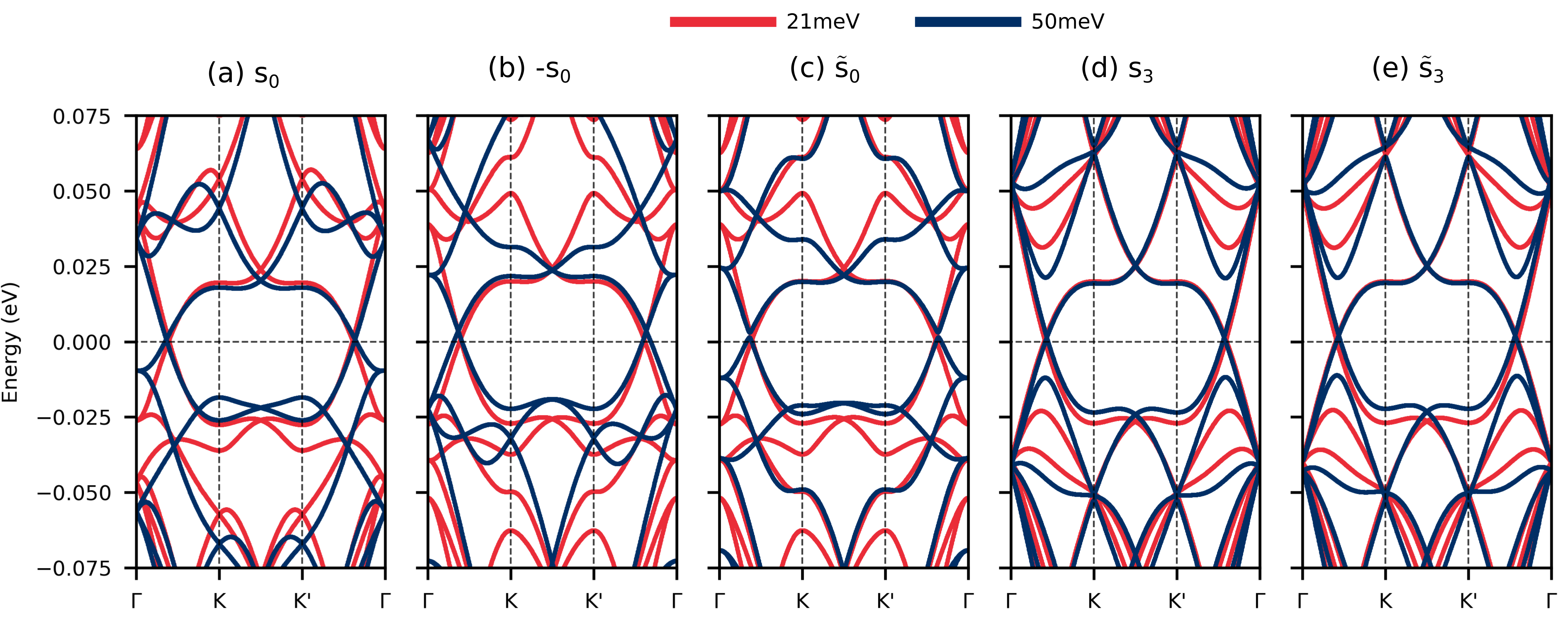}
    \caption{Band structure of an ABC TLG supercell upon which a periodic potential has been applied that corresponds to the different moir\'e perturbation terms of the 2-component model. The potential is only applied on the low-energy sites with an amplitude of 21~meV (red) and 50~meV (blue). (a) The applied potential is strictly even and of the form $\sum_i V_0 \cos (\mathbf{G}_i.\mathbf{r})$ and identical on both layers. (b) The same as (a) but with a negative sign. (c) The applied potential is of the form $\sum_i V_0 \sin (\mathbf{G}_i.\mathbf{r})$ and with opposite sign between both layers. (d) Potential of the form $\sum_i V_0 \cos (\mathbf{G}_i.\mathbf{r})$ that is identical on both layers. (e) The potential is of the form $\sum_i V_0 \sin (\mathbf{G}_i.\mathbf{r})$ and with opposite sign between both layers.}
    \label{fig:moire_bands_tlg}
\end{figure}

The same procedure is applied to an ABC-stacked TLG in the moir\'e unit cell for which the band structures are shown in figure~\ref{fig:moire_bands_tlg}. None of the contributions seems to open up a gap at the PDP or SDP, only a constant average mass term can open a gap at the PDP or perhaps the introduction of a intra- or interlayer pseudo-vector potential, which we do not consider here.

\begin{figure}[h]
    \centering
    \includegraphics{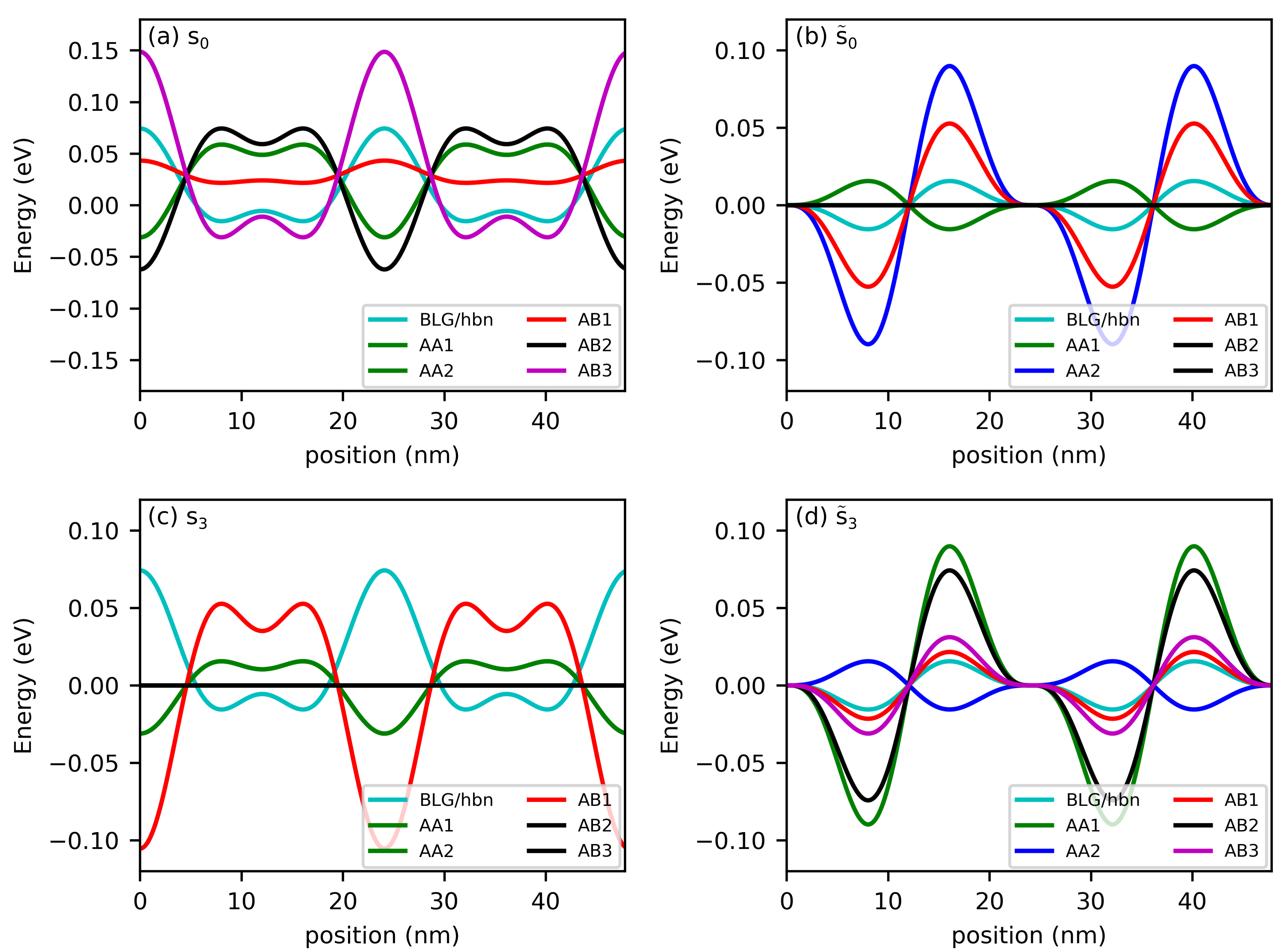}
    \caption{Cut along the $y$-axis of the different components of the 2D  moir\'e potential of the encapsulated BLG configurations for 2 times the period. The cut goes through the maxima of the potentials. Shown are the (a) $s_0$ term, (b) $\tilde{s}_0$ term, (c) $s_3$ term, and (d) $\tilde{s}_3$ term.}
    \label{fig:potential_cuts_blg}
\end{figure}

\begin{figure}[h]
    \centering
    \includegraphics{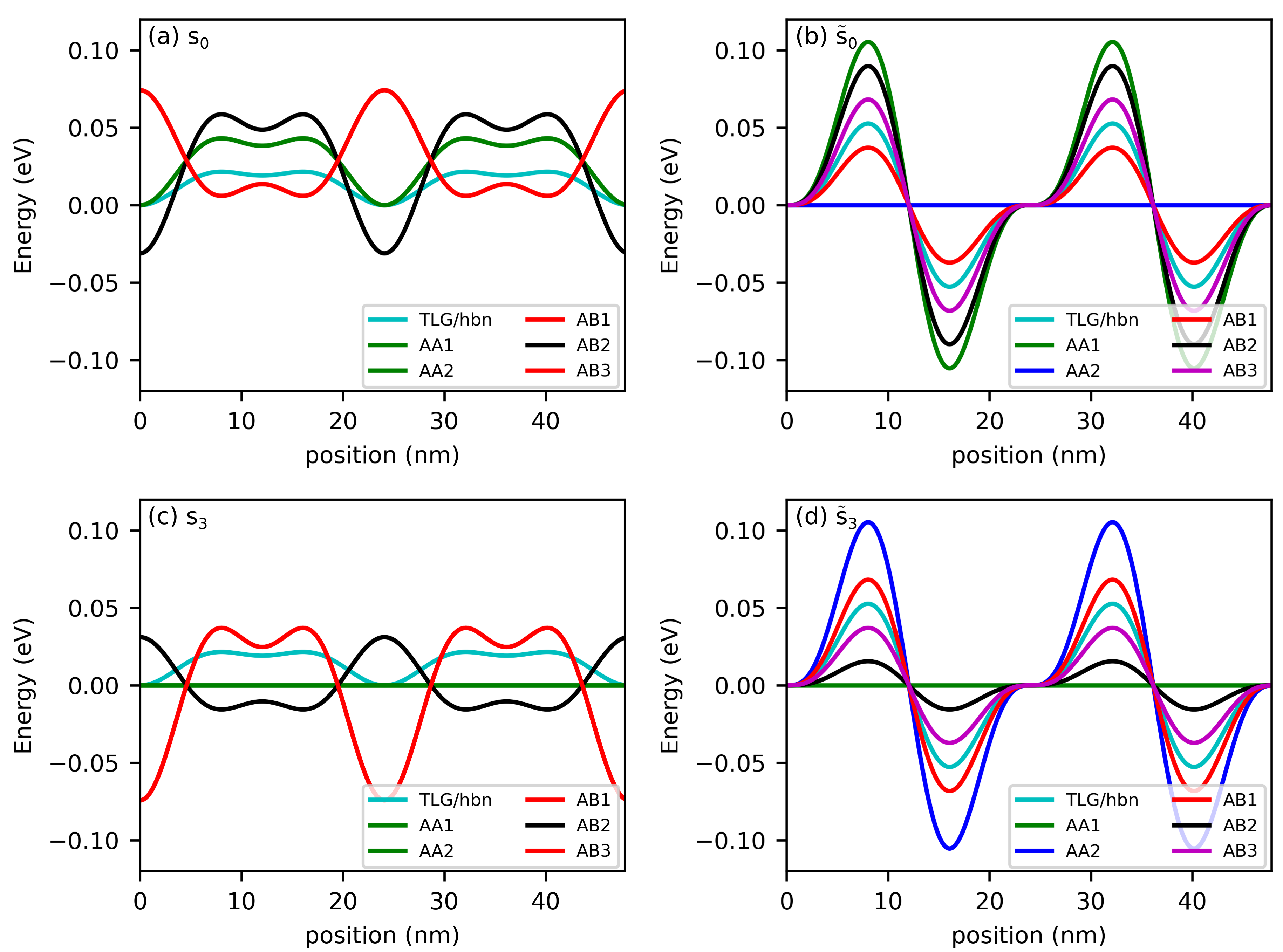}
    \caption{Cut along the $y$-axis of the different components of the 2D  moir\'e potential of the encapsulated TLG configurations for 2 times the period. The cut goes through the maxima of the potentials. Shown are the (a) $s_0$ term, (b) $\tilde{s}_0$ term, (c) $s_3$ term, and (d) $\tilde{s}_3$ term.}
    \label{fig:potential_cuts_tlg}
\end{figure}

In what follows, the terms in equation~\eqref{eq:scalar_spatial_moire} and \eqref{eq:mass_spatial_moire} are determined by setting up a continuum model for the graphene/hBN potential for each layer (see ref. \cite{moon_electronic_2014} for details) and applying the above procedure for the hBN encapsulated systems in the main text. A cut along the y-axis of the 2D moir\'e potentials are shown in figure~\ref{fig:potential_cuts_blg} for the BLG systems and figure~\ref{fig:potential_cuts_tlg} for the TLG systems. The AB1 BLG system, being the focus of the main text, shows a strong $s_3$ component, inducing a gap in the hole side and reducing the Fermi velocity near the SDP as is apparent from figure \ref{fig:moire_bands_blg} and readily observed in the fully atomistic result of the main text. The potentials for AB1 as described by equations \eqref{eq:scalar_spatial_moire} and \eqref{eq:mass_spatial_moire} and figure \ref{fig:potential_cuts_blg} are added to a BLG in a moiré unit cell for which we calculate the band structure and find a good agreement with the fully atomistic result. The BLG/hBN system also shows a large $s_3$ contribution which is observed when the band structure is calculated with the simplified model. unlike AB1 however, BLG/hBN shows no such feature in the atomistic result from the main text. Although a reasonable qualitative agreement is reached, some important features such as gaps at the SDP are not always consistently obtained when comparing the 2 methods. We conclude that the additional terms in equation \eqref{eq:moire_perturbation2} should be included in order to obtain an accurate analytical result. This model does however show the degree of tunability of the moiré pattern onto a multilayer graphene, simply by considering the precise positioning of aligned hBN layers.

\pagebreak

\section{TLG with different coupling to hBN}
In this section we perform the same atomistic tight-binding calculations as shown in Figure 2 in the main text for unrelaxed ABC-stacked trilayer graphene on hBN, but with a modified set of parameters describing the graphene/hBN interfaces. We compare graphene/hBN parameter sets as used in the BLG configurations by using the Slater-Koster type functions or the LDA parametrization. For the ABC stacked trilayer we use the LDA parameters shown in the main text. The relevant difference between these two sets is the value of  $\gamma_1$, 0.48eV for the Slater-Koster set versus 0.348eV for the LDA set. 

\begin{figure}
    \centering
    \includegraphics{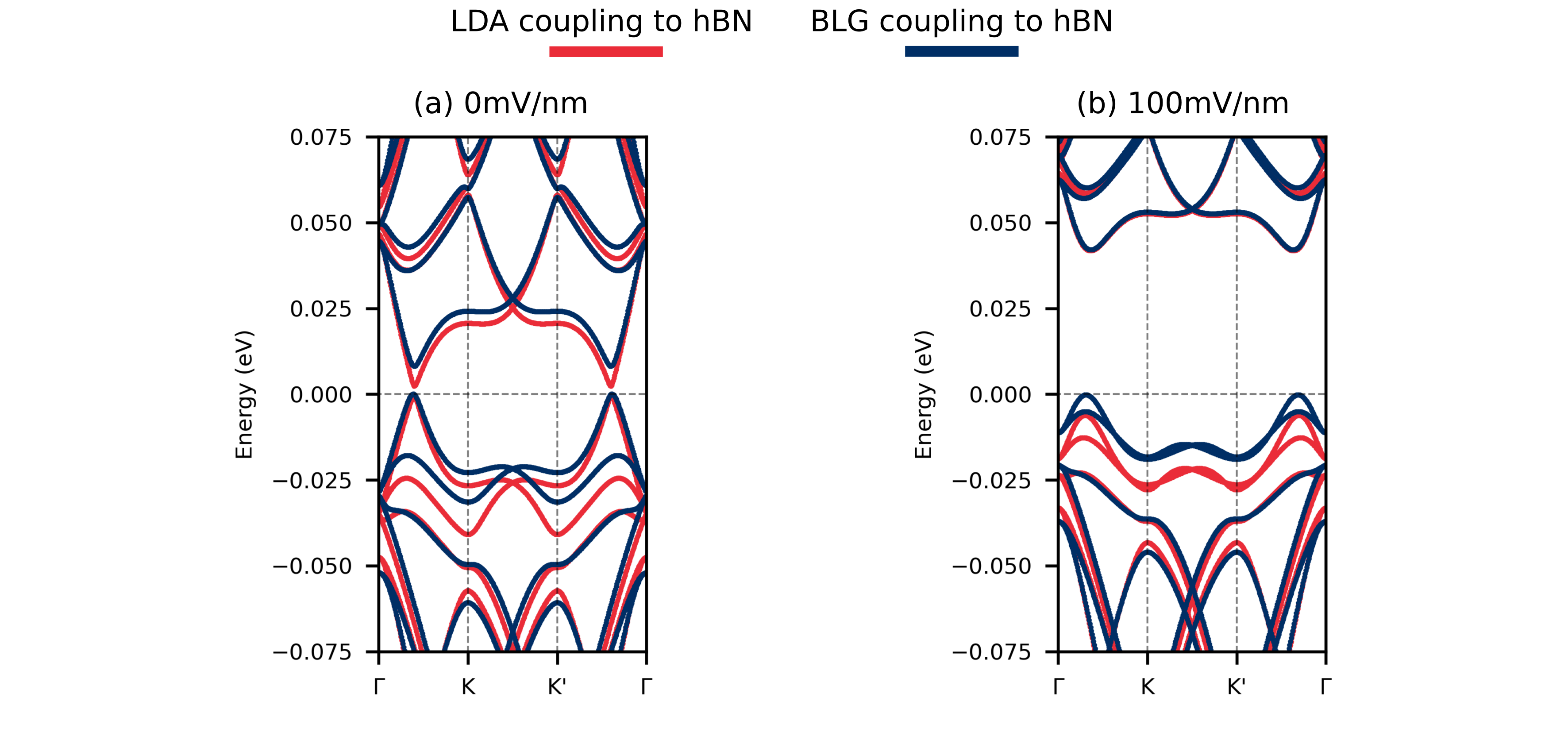}
    \caption{Electronic band-structures for the TLG/hBN system where a different parameter set is used for the hoppings between graphene and hBN and within the hBN layer. In red the LDA set is used, in blue the same parameter set as for the BLG configurations in the main text is used. (a) Has no applied electric field. (b) Has an applied electric field of 100mV/nm.}
    \label{fig:bands_tlg_hbn_comparison}
\end{figure}

The resulting electronic band-structures are shown in figure \ref{fig:bands_tlg_hbn_comparison} both with and without gating. The bands where the LDA parameters are used (red) show a smaller gap at both the PDP and SDP when no gating is applied and has a larger bandwidth in the plot with gating than the bands where the Slater-Koster parameters are used (blue).  From figure \ref{fig:bands_tlg_hbn_comparison} it is clear then that the occurrence of isolated flat bands depend on the exact parameters used for the coupling of graphene and hBN, especially when the effective gaps are small, which is the case here.

\clearpage
\section{Density of states as a function of gating}
\begin{figure}[h]
    \centering
    \includegraphics{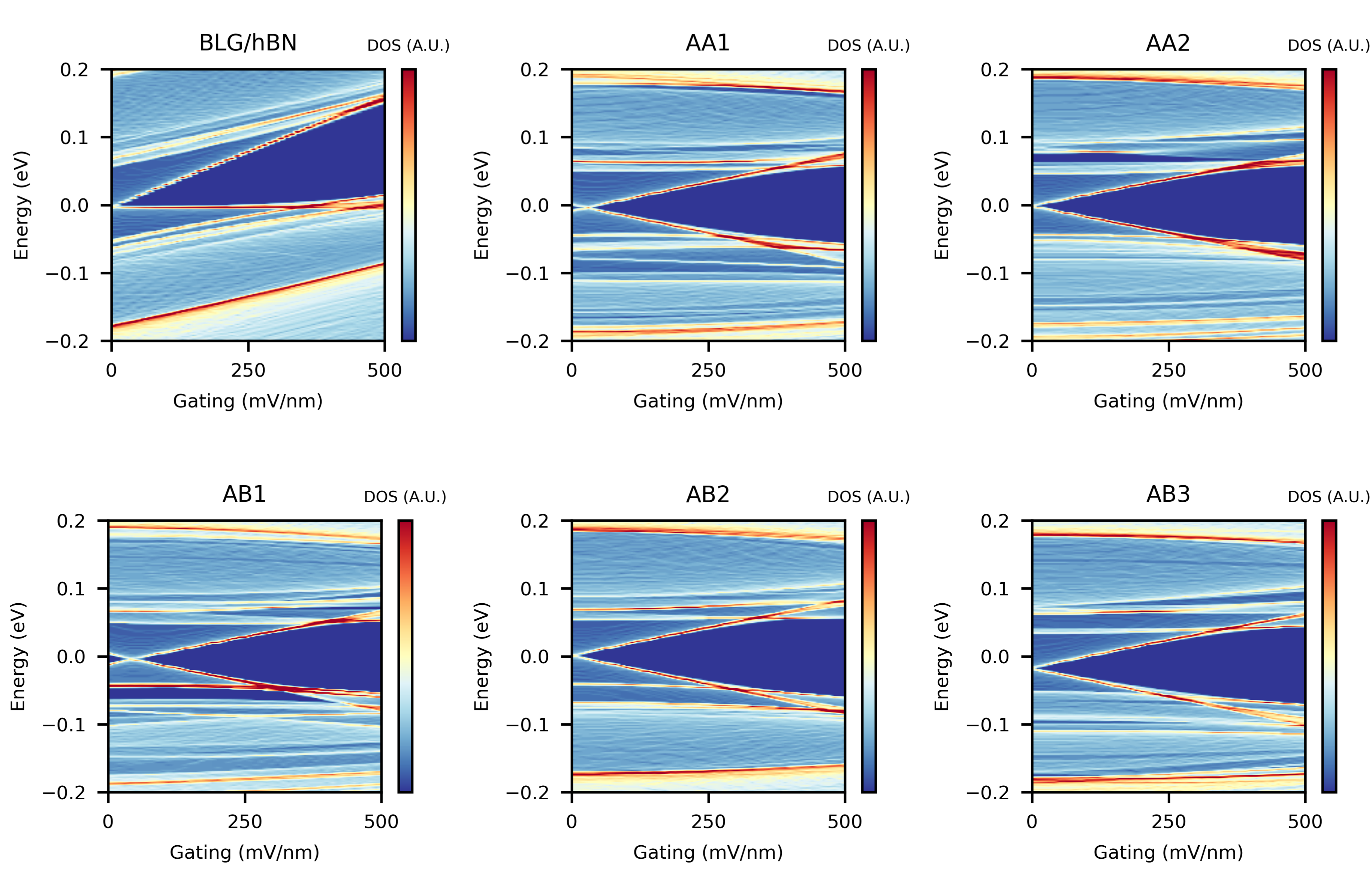}
    \caption{Dos as a function of gating at energies ranging from -0.2eV to 0.2eV for AB bilayer graphene systems. The DOS is taken along $\Gamma - K - K' - \Gamma$ and is given in arbitrary units.}
    \label{fig:sweeps_bilayer}
\end{figure}

\begin{figure}[h]
    \centering
    \includegraphics{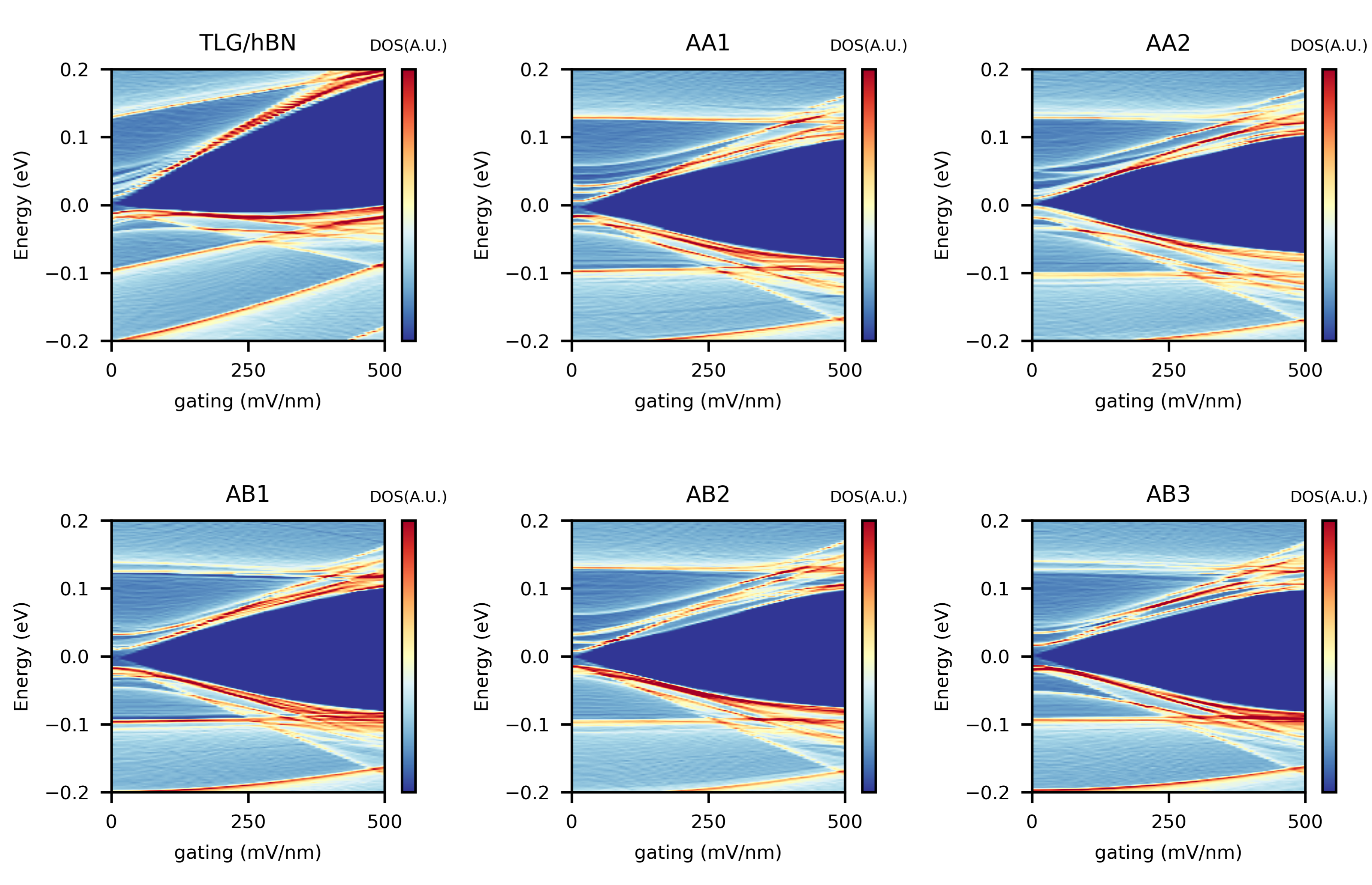}
    \caption{Dos as a function of gating at energies ranging from -0.2eV to 0.2eV for ABC trilayer graphene systems. The DOS is taken along $\Gamma - K - K' - \Gamma$ and is given in arbitrary units.}
    \label{fig:sweeps_ABC_trilayer}
\end{figure}

Here we present the density of states along the $\Gamma - K - K' - \Gamma$ path for all the BLG systems in figure~\ref{fig:sweeps_bilayer} and all the ABC TLG systems in figure~\ref{fig:sweeps_ABC_trilayer} both for which relaxation effects are present. The gating is implemented as described in the methods section in the main text. From these plots, gaps in the spectrum can be easily identified as well as their behaviour when gating is applied. The plots for graphene on a single hBN layer are skewed upwards due to the applied potential no longer being symmetric around zero when only considering the graphene layers.

\clearpage

\section{Average DOS per sublattice}
\begin{figure}[h]
    \centering
    \includegraphics{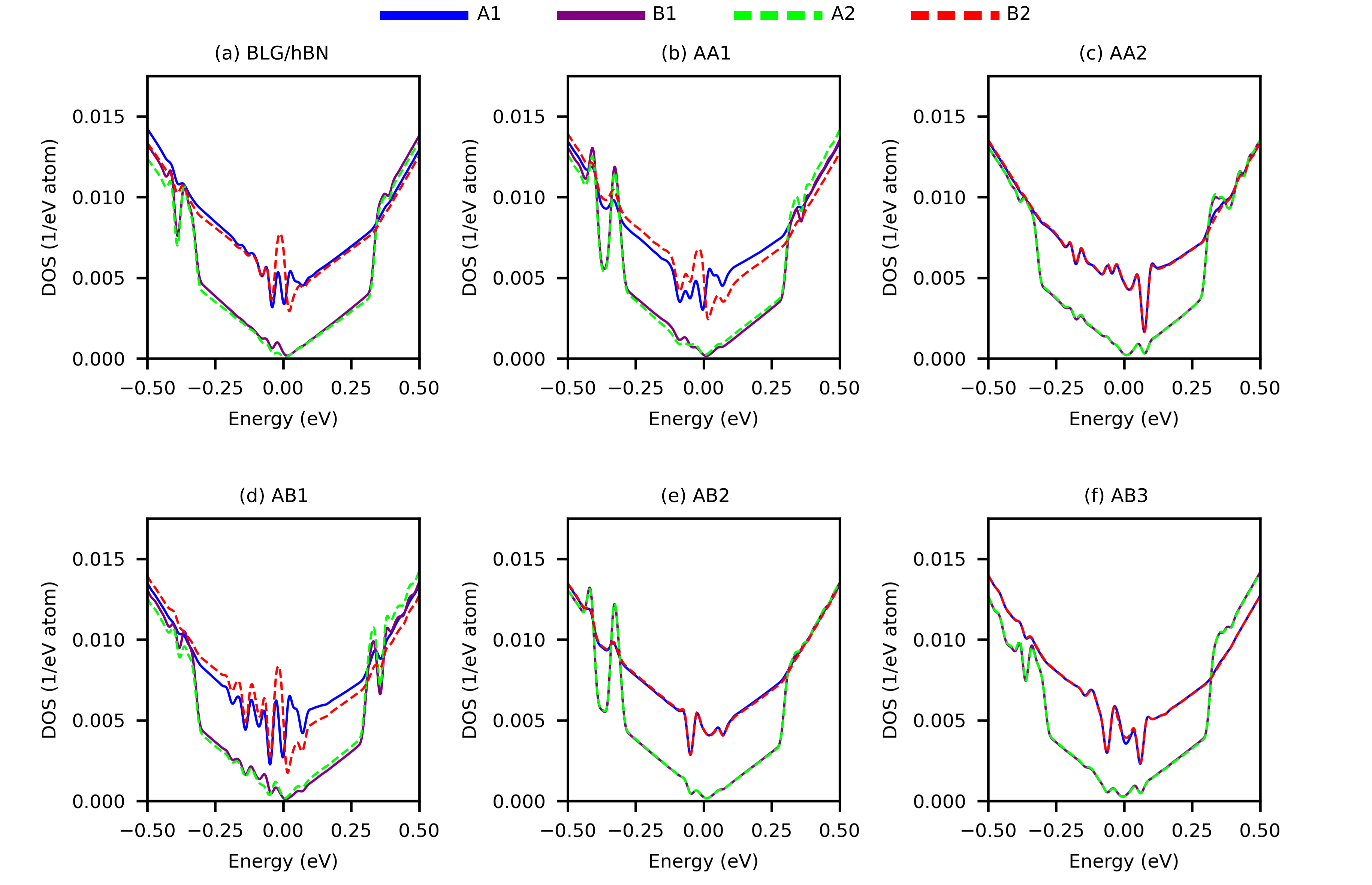}
    \caption{The average DOS per sublattice of bilayer graphene for the whole moir\'e supercell for the different configurations.}
    \label{fig:average_dos}
\end{figure}

In figure~\ref{fig:average_dos} the average DOS per sublattice of graphene is shown, where A1 and B2 are the low-energy (non-dimer) sites and B1 and A2 are the high-energy (dimer) sites. The different dips in the DOS correspond with the formation of the SDPs. It is also interesting to note that for systems with broken inversion symmetry, the layer degeneracy is broken as the DOS for the dimer and non-dimer sites are no longer identical.

\clearpage

\section{In-plane strain}
\begin{figure}
    \centering
    \includegraphics{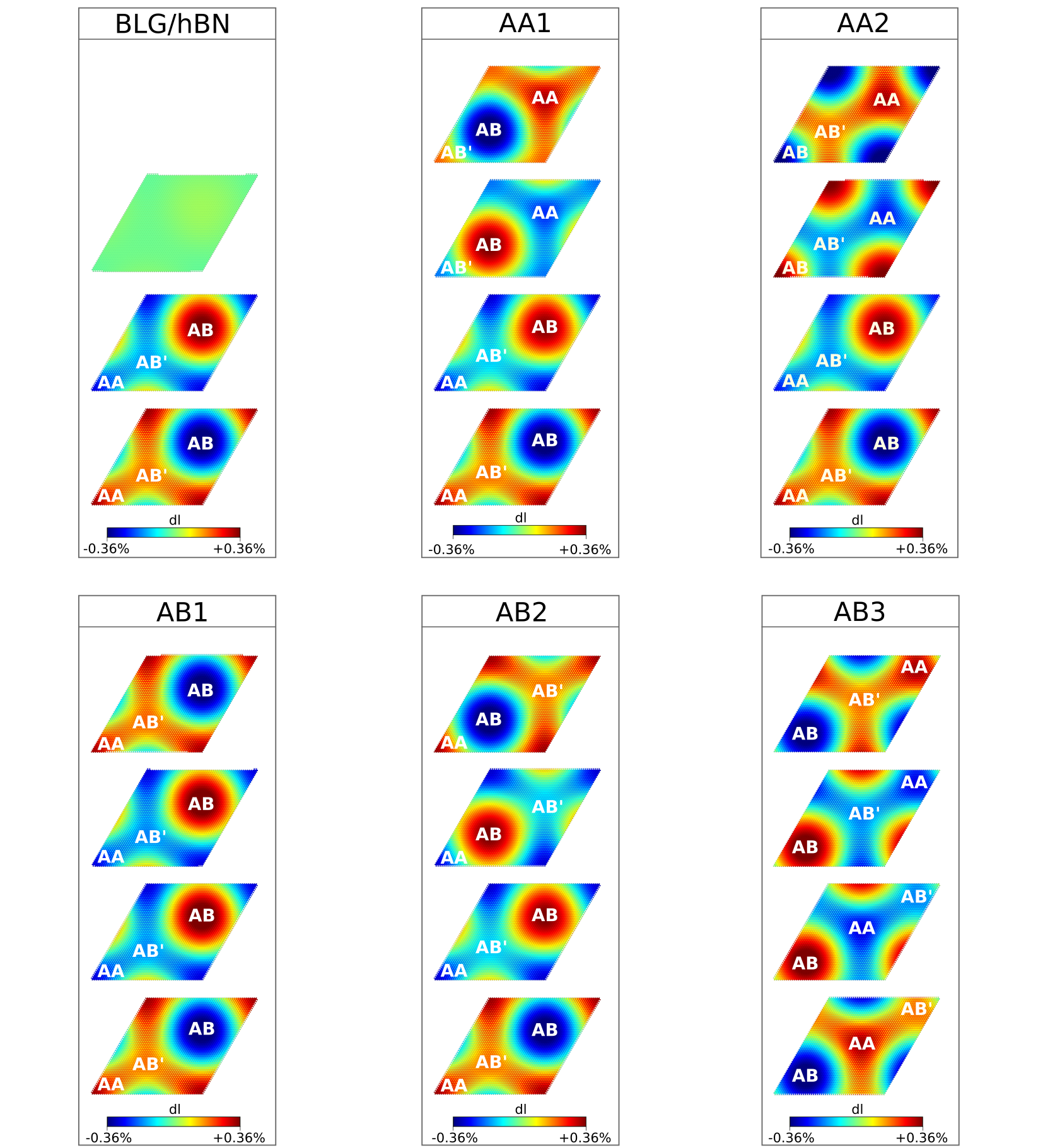}
    \caption{Maps for the BLG systems indicating the in-plane strain due to relaxation effects as a percentage of the original bond length, which are $a_{gr}$ 0.142nm for graphene and $a_{hBN}$ = 1.018 $a_{gr}$ for hBN.}
    \label{fig:strain_maps_blg}
\end{figure}

\begin{figure}
    \centering
    \includegraphics{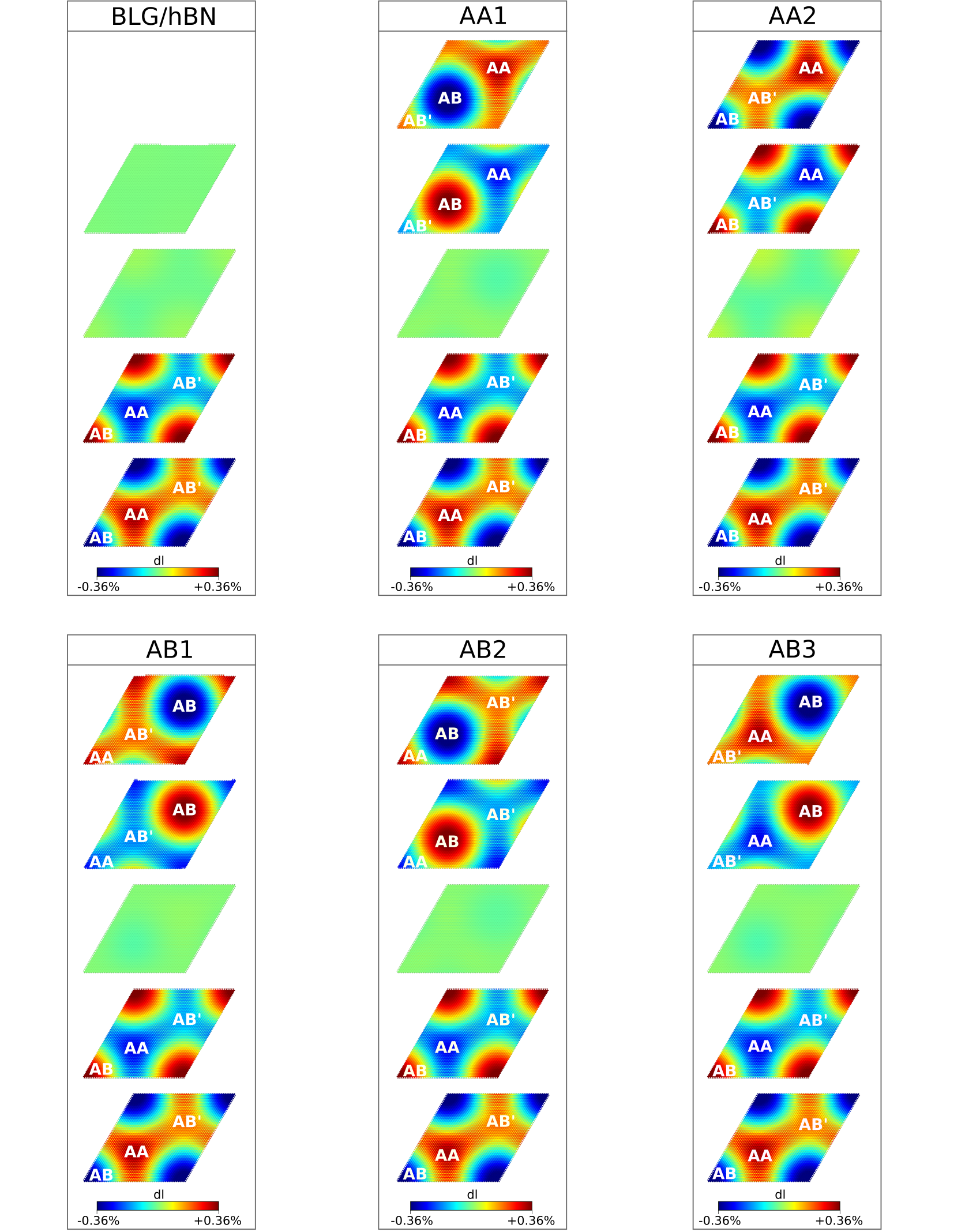}
    \caption{Maps for the TLG systems indicating the in-plane strain due to relaxation effects as a percentage of the original bond length, which are $a_{gr}$ 0.142nm for graphene and $a_{hBN}$ = 1.018 $a_{gr}$ for hBN.}
    \label{fig:strain_maps_tlg}
\end{figure}

In this section we provide maps of the in-plane strain that is present in the different systems. The local stacking regions are indicated in each map. Three distinct local stacking configurations can be defined, AA which has both boron and nitrogen on top of a carbon atom, AB which has boron on top of carbon and nitrogen in the centre of a carbon ring and AB' which has nitrogen on top of carbon with boron in the centre of a carbon ring. As mentioned in the main text, the AB stacked regions are expanded in the graphene layer and contracted in the hBN layer due to it being the most energetically favourable stacking configuration, this way the area containing this configuration is maximized. The strain is accumulated in surrounding regions in the shape of domain walls with AA and AB' stacked regions \cite{argentero_unraveling_2017}. As is clear from figures~\ref{fig:strain_maps_blg} and \ref{fig:strain_maps_tlg}, the in-plane strain is almost entirely determined by the local stacking between graphene and hBN and the in-plane strain in neighbouring graphene layers do not affect each other. This is also clearly visible since the middle graphene layer in the TLG systems has strongly reduced in-plane strain as it neighbours no hBN layer.

\clearpage

\clearpage

\section{Pseudo-magnetic field due to in-plane strain}
\begin{figure}[h]
    \centering
    \includegraphics{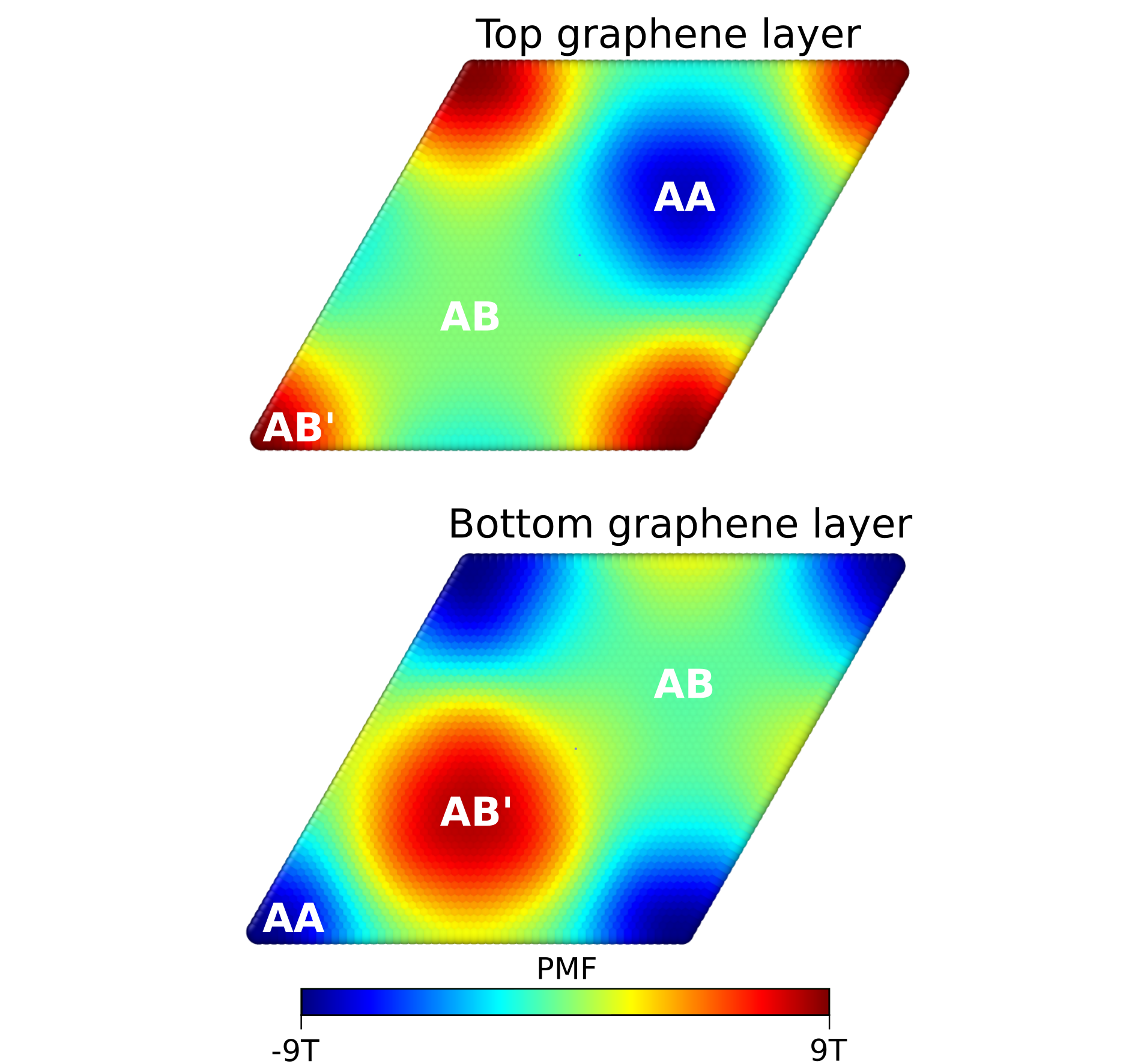}
    \caption{The pseudo-magnetic field on the low-energy sites in the graphene layers for the AA1 BLG system as a result of in-plane strain.}
    \label{fig:pmf}
\end{figure}

Here we show the pseudo-magnetic field (PMF) that emerges due to the in-plane strain caused by relaxation effects. The values of the PMF for all systems are on the order of 9~T, being different than the 40~T which was found in Ref. \cite{jung_moire_2017}. Here the PMF is entirely dependent on the local stacking between graphene and hBN so that the PMF for all systems can be further extrapolated from figure~\ref{fig:pmf}.

\clearpage
\newcommand{\newblock}{}
\bibliographystyle{iopart-num}
\bibliography{My_Library_modified}